\documentclass[a4paper,10pt]{amsart}
\usepackage{fancyhdr}
\usepackage{natbib}
\usepackage{booktabs}
\usepackage{graphicx}
\usepackage{hyperref}

\begin{document}

\title[An analytical solution of the Euler equations over orography]{An analytical solution of the stationary fully-compressible linear Euler equations over orography}

\author{Juan Simarro}
\author{Petra Smolikova}
\author{Jozef Vivoda}
%\address{AEMET, c/Botanico Cavanilles 3, Valencia 46010, Spain}
%\address{Czech Hydrometeorological Institute, Prague, Czech Republic}
%\address{Slovak Hydrometeorological Institute, Bratislava, Slovakia}

\begin{abstract}
An analytical linear solution of the fully compressible Euler equations is found, in the particular case of a stationary two dimensional flow that passes over an orographic feature with small height-width ratio. A method based on the covariant formulation of the Euler equations is used, and the analytical vertical velocity as well as the horizontal velocity, density and pressure, are obtained. The analytical solution is tested against a numerical model in three different regimes, hydrostatic, non-hydrostatic and potential flow. The model used is a non-hydrostatic spectral semi-implicit model, with a height-based vertical coordinate. It is shown that there is a clear and consistent convergence of the numerical solution towards the analytical solution, when the resolution increases. The method described is intended to be used as an idealized test for numerical weather models.
\end{abstract}

\maketitle

%\keywords{Euler equations; analytical solution; numerical model; idealized test; gravity waves}

%\footnotetext[2]{}

% ===============================================================================

\section{Introduction}

% Nowadays, high resolution weather numerical models are based on the fully compressible Euler equations. For instance, ALADIN \citep{bubnova1993}, COSMO \citep{baldauf2007}, WRF \citep{skamarock2005} and UM \citep{davies2005} models, just to cite a few of them. 

% The shallow water equations are commonly used to test those schemes related to the horizontal discretization. The suite of tests detailed in \cite{williamson1992} includes advection across the poles, steady state geostrophically balanced flow, forced nonlinear advection of an isolated low, zonal flow impinging on an isolated mountain and Rossby-Haurwitz waves. These are two dimensional horizontal tests which, thereby, are not suited for testing the vertical discretization. 

During the design and development of a weather numerical model, it is customary to perform a number of tests. The tests found in the literature are suited for different purposes, and there are specific tests for examining vertical discretization schemes. The vertical slice test designed by \cite{skamarock1994}, which consists in non-stationary gravity waves, is used for checking the spatial and time discretization schemes in a linear regime without orography. Another vertical slice test, described in \cite{straka1993}, consists in a density current, and serves to check non-linear non-hydrostatic regimes with diffusion. The vertical slice tests in \cite{bryan2002} include condensation and evaporation processes, and simplified cloud microphysics. Other interesting vertical slice tests are those found in \cite{schaer2002}, \cite{klemp2003} and \cite{girad2005}.

We are interested in vertical slice tests for checking a stationary flow over an orographic feature with small height-width ratio, where a set of gravity waves are generated. Usually, this kind of tests are performed and checked against analytical solutions that come from a simplified version of the Euler equations, for instance, the Boussinesq equations. 

Our intention is to find an analytical solution of the fully compressible Euler equations, in the particular case of a stationary flow that passes over an arbitrarily shaped smooth orography. This problem has been studied \citep{smith1979,lin2007,laprise1989}. However, in this work we apply a different method for finding the analytical solution, and we check its consistency using a non-linear non-hydrostatic numerical model.

The starting point of the method proposed in this paper is a spatial domain delimited by a flat orography, where a stationary and trivial solution is known. This solution is an isothermal and stratified flow, with constant horizontal velocity. This is the background state, which includes a flat orography in its definition. A second stationary state is taken into account, the perturbed state, for which the domain is deformed in the lower limit with an arbitrarily shaped orography. In doing so, a set of stationary waves appears in the flow. The definition of the perturbed state includes not only the stationary waves, but also the deformation of the lower limit of the domain by the orography.

The way chosen for solving the stationary Euler equations in a domain deformed by a orography is to write the Euler equations in a coordinate independent formulation, also called covariant formulation, as dictated by the differential geometry \citep{aubin2001}. In this formulation, the free slip condition in the lower part of the domain is simplified to set the vertical component of the velocity to zero at the surface. That is, the orography is not involved in the lower boundary condition, which is reduced to a trivial condition, and it only appears explicitly in the equations. The idea is that the orography is moved from the boundary conditions into the equations, and this change is achieved using the covariant formulation. We provide the details of this procedure in section \ref{analyticalsolution}.

Then, after writing the Euler equations in a covariant form with an arbitrarily shaped orography, the equations are linearized. The linearization procedure includes, not only the velocity components, density and pressure perturbations, but the orographic terms that appears in the covariant version of the Euler equations. It comes into view that the orographic terms act as forcing terms for the momentum equations, and are responsible for the waves generated in the flow. Then, the linearized equations are solved, and the perturbations of the velocity components, density and pressure are found.

The analytical solution of the Euler equations for a stationary flow that passes over a hill, following the method outlined in the previous paragraphs, is described in Section \ref{analyticalsolution}. In section \ref{modelsimulations} the numerical simulations are exposed and compared to the analytical solutions. Three types of flow are considered, hydrostatic, non-hydrostatic and potential. Finally, the conclusions are pointed out in section \ref{conclusions}.

\section{Analytical solution}
\label{analyticalsolution}

% GEOMETRY
We consider a two dimensional channel of length $L$. In the vertical dimension, the upper boundary is open and the lower boundary is limited by the orography, $B(x)$. 

% EQUATIONS
The Euler equations are

\begin{align}
\frac{d u}{d t} + \frac{1}{\rho}\frac{\partial p}{\partial x} & = 0, \\
\frac{d w}{d t} + \frac{1}{\rho}\frac{\partial p}{\partial z} + g & = 0, \\
\label{201326}
\frac{\partial \rho}{\partial t} + \nabla ({\bf u} \rho) & = 0, \\
\label{201345}
\frac{d p}{d t} -c_s^2 \frac{d \rho}{d t} & = 0,
\end{align}
where ${\bf u} = (u, w)$ is the velocity vector, $\rho$ the density, $p$ the pressure, $g$ the acceleration due to gravity, and $c_s^2$ the speed of sound, given by

\begin{align*}
c_s & = \sqrt{\frac{c_p}{c_v} RT}, \\
T & = \frac{p}{R \rho}.
\end{align*}

As we consider stationary states, the partial time derivative vanishes. The total time derivative is reduced to the advective part, that is

\begin{align*}
\frac{d}{dt} = {\bf{u}} \cdot \nabla.
\end{align*}
%

% BOUNDARY CONDITIONS
The boundary conditions imposed to the solution are the following. The free slip boundary condition is used at the bottom, meaning that there is no flux of mass through that surface. It is

\begin{align}
\label{200904}
w(x,z=B(x)) = u(x,z=B(x)) \, \frac{\partial B}{\partial x}.
\end{align}

In the upper limit, considered to be at the infinity, density, and therefore pressure, tends to vanish. The mathematical condition for the density is then

\begin{align*}
\lim_{z \to +\infty} \rho(z) = 0.
\end{align*}

On the other hand, the lateral boundaries are periodic, and consequently we impose the condition $\psi(0, z) = \psi(L, z)$ for any function $\psi(x,z)$ involved in the problem. As it is shown later, this periodicity will permit to solve the problem by Fourier transforming the equations in the horizontal dimension.

% COORDINATE TRANFORMATION

The boundary condition for the velocity at the lower limit, given by (\ref{200904}), involves both the horizontal and vertical components of the velocity, as well as the orography, in a non linear way. Imposing this condition can be cumbersome. We propose a method based on a coordinate transformation to circumvent this problem, so that the boundary condition is transformed into a trivial condition in the new coordinate system.

The proposed change of coordinates is simple. The new coordinates, named $(X,Z)$ are related to the original euclidean coordinates $(x,z)$ by

\begin{align}
\label{211208}
x & = X, \\
\label{211209}
z & = Z + B(X),
\end{align}
where $B(X)$ is the orography. With this change of coordinates, the lower limit is the coordinate line $Z=0$. The Jacobian of this transformation is

\begin{align*}
J = \frac{\partial (x,z)}{\partial (X,Z)} =
\left( \begin{array}{ccc}
1 & 0 \\
B_X & 1 \end{array} \right),
\end{align*}
where the subindex in $B_X$ means partial derivative with respect to coordinate $X$. The contravariant components of the velocity in the original and new coordinates, named respectively $(u,w)$ and $(U,W)$, are related by

\begin{align}
\label{201419}
u & = U, \\
\label{201420}
w & = B_X U + W,
\end{align}
which implies that the boundary condition for the velocity given by (\ref{200904}) is reduced to a very simple and convenient condition, which is

\begin{align}
\label{201418}
W(X,Z=0) = 0.
\end{align}

The relations (\ref{201419}) and (\ref{201420}), are valid for any contravariant vector, not only for the velocity. From them, we observe that the acceleration due to gravity, the contravariant vector $(0, g)$, has the same components in the new coordinate system.

The Euler equations must be transformed into the new coordinate system. To this end, the metric tensor must be found, as well as the Christoffel symbols. The euclidean metric, $\eta_{xz}$, is the identity in the original coordinate system, whereas in the new coordinate system it is

\begin{align*}
\eta_{XZ} = 
\left( \begin{array}{ccc}
1 + B_X^2 & B_X \\
B_X & 1 \end{array} \right).
\end{align*}

The determinant is $|\eta_{XZ}| = 1$, and the inverse of the metric tensor is

\begin{align}
\label{201318}
\eta^{-1}_{XZ} = 
\left( \begin{array}{ccc}
1 + B_X^2 & -B_X \\
-B_X & 1 \end{array} \right).
\end{align}

Finally, we need the Christoffel symbols for the advection term of the contravariant components of the velocity. The only non-zero symbol is

\begin{align}
\label{201229}
\Gamma^Z_{XX} = B_{XX}.
\end{align}

The advection of a contravariant vector $\bf V$ by the velocity field $\bf U$ is

\begin{align}
\label{201228}
\left({\bf U} \cdot \nabla) \, {\bf V} \right)^i 
= U^j \frac{\partial}{\partial X^j} V^i + \Gamma^i_{jk} U^j U^k,
\end{align}
where the indexes, running over $1$ and $2$, are referencing the coordinates $X$ and $Z$ respectively (that is, $X^1 = X$, $X^2 = Z$, and $U^1 = U$, $U^2 = W$). Then, from (\ref{201229}) and (\ref{201228}) the advection of the contravariant velocity components by the velocity itself is

\begin{align*}
({\bf U} \cdot \nabla) \, U  & = (U \, \frac{\partial}{\partial X} + W \, \frac{\partial}{\partial Z}) \, U , \\
({\bf U} \cdot \nabla) \, W  & = (U \, \frac{\partial}{\partial X} + W \, \frac{\partial}{\partial Z}) \, W + B_{XX} \, U^2,
\end{align*}
where the new term $B_{XX} \, U^2$, appearing in the advection of the vertical component of the velocity, is due to the non-zero Christoffel symbol (\ref{201229}). It is usual to interpret those terms as inertial forces.

The pressure gradient is

\begin{align*}
(\nabla p)^i = (\eta^{ij} \, \frac{\partial}{\partial X^j}) \, p,
\end{align*}
and then, from the inverse of the metric tensor given in (\ref{201318}), the contravariant pressure gradient writes

\begin{align*}
(\nabla p)^X & = ((1 + B_X^2) \frac{\partial}{\partial X} - B_X \frac{\partial}{\partial Z}) \, p, \\
(\nabla p)^Z & = (-B_X \frac{\partial}{\partial X} + \frac{\partial}{\partial Z}) \, p.
\end{align*}

The divergence term of the continuity equation (\ref{201326}) is written in the new coordinate system as

\begin{align*}
\nabla \cdot (\rho {\bf{U}}) = 
\frac{1}{\sqrt{|g|}} \, \frac{\partial}{\partial X^i} ( \sqrt{|g|} \, \rho \, U^i),
\end{align*}
and, taking into account that $|g| = 1$ the divergence remains written in the same form as in the euclidean coordinate system, that is

\begin{align*}
\nabla \cdot (\rho {\bf{U}}) = \frac{\partial}{\partial X} \left( \rho \, U \right) + \frac{\partial}{\partial Z} \left( \rho \, W \right).
\end{align*}

Finally, the advection of the pressure and density in the equation (\ref{201345}), as well as the advection of any scalar function $\psi$, is simply

\begin{align*}
\left( {\bf{U}} \cdot\nabla \right) \, \psi
= ( U \frac{\partial}{\partial X} + W \frac{\partial}{\partial Z}) \, \psi.
\end{align*}

Using the previous results, the non linear stationary Euler equations are written  in the new coordinate system $(X,Z)$ as

\begin{align}
\label{231812}
\frac{DU}{Dt}   + \frac{1}{\rho} \,
((1 + B_X^2) \frac{\partial p}{\partial X} - B_X \frac{\partial p}{\partial Z} ) & = 0, \\
\label{231813}
\frac{DW}{Dt} + B_{XX} \, U^2 +
\frac{1}{\rho} \, (-B_X \frac{\partial p}{\partial X} + \frac{\partial p}{\partial Z}) + g & = 0, \\
\label{231814}
\frac{\partial}{\partial X} \left( \rho \, U \right) + 
\frac{\partial}{\partial Z} \left( \rho \, W \right) & = 0, \\
\label{231815}
\frac{Dp}{Dt} - c_s^2 \, \frac{D \rho}{Dt} & = 0,
\end{align}
where

\begin{align*}
\frac{D \psi}{Dt} \equiv U \, \frac{\partial \psi}{\partial X} + W \, \frac{\partial \psi}{\partial Z}.
\end{align*}

At this point, we stress that the Euler equations for the horizontal and vertical momentum (\ref{231812}) and (\ref{231813}) contains orographic terms, that is, terms with the orography function $B(X)$, whereas the lower boundary condition is independent of the orographic features, reduced to be

\begin{align*}
W(X,Z=0)=0.
\end{align*}

This is why we can say that, by writing the equations in a covariant form, we move the orography $B(X)$ from the boundary condition into the Euler equations. 

% LINEAR
As has been already mentioned, we are interested in finding out a linear solution of the stationary Euler equations. We have been inspired by the methods used in \cite{baldauf2013}, were the authors found a non stationary solution of the linear Euler equations in a two dimensional atmosphere, with a flat lower boundary and an upper boundary placed at a fixed and finite height. In this work, instead, we look for a stationary solution, with a non flat bottom boundary and without an upper boundary. 

% Our solution is less general in the aspect that it is stationary, although more general if we focus on the inclusion of non flat orography.

As already said, we consider an arbitrary orographic obstacle of small amplitude given by the function $B(X)$. That is, because $B(X)$ is small, we expect that the waves that are produced in a flow that passes over this obstacle will be of small amplitude. In the linearization procedure, we will only retain the orographic terms that are at most first order, rejecting higher order terms. 

We rewrite the Euler equations (\ref{231812}) to (\ref{231815}), placing in the right hand side the linear orographic terms, which will be treated as forcing terms of the equations

\begin{align}
\label{231831}
\frac{DU}{Dt}  + \frac{1}{\rho} \, \frac{\partial p}{\partial X} & = 
\frac{B_X}{\rho} \, \frac{\partial p}{\partial Z}, \\
\label{231832}
\frac{DW}{Dt} + \frac{1}{\rho} \, \frac{\partial p}{\partial Z} + g & = \frac{B_X}{\rho} \, \frac{\partial p}{\partial X} -B_{XX} \, U^2 , \\
\label{231833}
\frac{\partial}{\partial X} \left( \rho \, U \right) + 
\frac{\partial}{\partial Z} \left( \rho \, W \right) & = 0, \\
\label{231834}
\frac{Dp}{Dt} - c_s^2 \, \frac{D \rho}{Dt} & = 0.
\end{align}

Linearising the Euler equations implies the existence of a steady reference state, being the solution a small perturbation around it. The reference unperturbed state is a steady hydrostatic solution of the Euler equations (\ref{231831}) to (\ref{231834}), with a flat lower boundary, that is $B(X) = 0$. Moreover, the reference state is isothermal ($T_0$) with a constant horizontal velocity ($U_0$), and it is completely determined if the surface pressure is given ($p_s$). Then, the background state has a flat orography $B(X)=0$, a constant velocity field defined by the contravariant components $(U_0, 0)$, and a pressure and density distribution which depends on the vertical coordinate $Z$ through

\begin{align}
\label{211206}
p_0(Z) & = p_s \, e^{-\delta Z}, \\
\label{211207}
\rho_0(Z) & = \rho_s \, e^{-\delta Z},
\end{align}
where

\begin{align*}
\rho_s & \equiv \frac{p_s \delta}{g}, \\
\delta & \equiv \frac{g}{R T_0}.
\end{align*}

For later use, we mention that the Brunt-V\"ais\"al\"a frequency of the background state is constant, equal to

\begin{align*}
N_0 = \frac{g}{\sqrt{c_p T_0}}.
\end{align*}

The linear version of the Euler equations (\ref{231831}) to (\ref{231834}) are found to be

\begin{align}
\label{231921}
U_0 \frac{\partial U'}{\partial X} + \frac{1}{\rho_0} \frac{\partial p'}{\partial X} & = 
-B_X \, g, \\
\label{231922}
U_0 \frac{\partial W'}{\partial X} + \frac{g}{\rho_0} \, \rho' + \frac{1}{\rho_0} \frac{\partial p'}{\partial Z} & = 
-B_{XX} \, U_0^2, \\
\label{231923}
\frac{\partial U'}{\partial X} + (\frac{\partial}{\partial Z} - \delta) \, W' +  \frac{U_0}{\rho_0} \frac{\partial \rho'}{\partial X} & = 0, \\
\label{231924}
\delta ( p_0 - c_s^2 \rho_0) \, W' + U_0 \frac{\partial}{\partial X} \, (c_s^2 \, \rho'- p') & = 0,
\end{align}
where the perturbed quantities are $\psi' = \psi - \psi_0$. Observe that the horizontal and vertical momentum equations have orographic forcing terms, whereas the continuity and the thermodynamic equations are free of them. In order to solve this linear system, following \cite{baldauf2013}, we apply the \cite{bretherton1966} transformation to the linear system (\ref{231921}) to (\ref{231924}). In doing so, as we show below, we obtain a new liner system where the coefficients for the variables are constants. The perturbations $\psi'$ are transformed to the Bretherton variables $\hat{\psi}$ in this way

\begin{align}
\psi'(X,Z) & \equiv \gamma^\pm(Z) \, \hat{\psi}(X,Z), \\
\gamma^\pm(Z) & \equiv  e^{\pm \frac{\delta}{2} Z},
\end{align}
being $U' = \gamma^+ \, \hat{U}$ and $W' = \gamma^+ \, \hat{W}$, whereas $p' = \gamma^- \, \hat{p}$ and $\rho' = \gamma^- \, \hat{\rho}$. The linear system (\ref{231921}) to (\ref{231924}) is written, in terms of the Bretherton variables, as

\begin{align}
\label{241756}
U_0 \frac{\partial \hat{U}}{\partial X} + \frac{1}{\rho_s} \frac{\partial \hat{p}}{\partial X} & = 
-B_X \, g \, \gamma^-, \\
\label{241757}
U_0 \frac{\partial \hat{W}}{\partial X} + \frac{g}{\rho_s} \, \hat{\rho} + \frac{1}{\rho_s} ( \frac{\partial}{\partial Z} - \frac{\delta}{2}) \, \hat{p}  & = 
-B_{XX} \, U_0^2 \, \gamma^-, \\
\label{241758}
\frac{\partial \hat{U}}{\partial X} + (\frac{\partial}{\partial Z} - \delta) \, \hat{W} +  \frac{U_0}{\rho_s} \frac{\partial \hat{\rho}}{\partial X} & = 0, \\
\label{241759}
\delta \rho_s (\frac{g}{\delta} - c_s^2) \, \hat{W} - U_0 \frac{\partial}{\partial X} \, (\hat{p} - c_s^2 \, \hat{\rho}) & = 0.
\end{align}

In the case of a finite horizontal domain of length $L$, the variables can be expanded in Fourier series (a Fourier transformation would be used for an infinite horizontal domain). Then, any variable $\hat{\psi}$ is transformed to $\tilde{\psi}$ following

\begin{align*}
\hat{\psi}(X,Z) = \sum_{k \in \frac{2\pi}{L} \mathbb{Z}} \tilde{\psi}(k,Z) \, e^{ikX}.
\end{align*}

Finally, the linear system to solve given in equations (\ref{241756}) to (\ref{241759}) is

\begin{align}
\label{250001}
i k U_0 \, \tilde{U} + \frac{ik}{\rho_s} \tilde{p} & = 
-g \, ik \tilde{B}(k) \, \gamma^-(Z), \\
\label{250002}
ik U_0 \, \tilde{W} + \frac{g}{\rho_s} \, \tilde{\rho} + \frac{1}{\rho_s} ( \frac{\partial}{\partial Z} - \frac{\delta}{2}) \, \tilde{p}  & = U_0^2 \, k^2 \tilde{B}(k) \, \gamma^-(Z), \\
\label{250003}
ik \tilde{U} + (\frac{\partial}{\partial Z} - \delta) \, \tilde{W} +  \frac{ik U_0}{\rho_s} \tilde{\rho} & = 0, \\
\label{250004}
\delta \rho_s (\frac{g}{\delta} - c_s^2) \, \tilde{W} - ik U_0 \, (\tilde{p} - c_s^2 \, \tilde{\rho}) & = 0,
\end{align}
where $\tilde{B}(k)$ are the Fourier coefficients of the orography function $B(X)$. The linear system (\ref{250001}) to (\ref{250004}), with a vertical derivative operator, is solved in the following way. The system is manipulated in order get to an equation where the only variable is the vertical component of the velocity. The result is a second order differential equation, which is solved to obtain

\begin{align}
\label{291410}
\tilde{W}(k,Z) & = ik U_0 \tilde{B}(k) (e^{\beta Z} - e^{-\frac{\delta}{2} Z}),
\end{align}
where

\begin{align}
\label{250156}
\beta^2 & \equiv {k^2 \alpha_0 - \frac{N_0^2}{U_0^2} + \frac{\delta^2}{4}}, \\
\alpha_0 & \equiv 1 - \frac{U_0^2}{c_s^2}.
\end{align}

We observe, from (\ref{250156}), that $\beta$ can be a real or an imaginary number, depending on the value of the wave number $k$. For values of $k$ such that 

\begin{align*}
k^2 < \frac{1}{\alpha_0} (\frac{N_0^2}{U_0^2} - \frac{\delta^2}{4}),
\end{align*}
$\beta^2$ is a negative real number, and therefore $\beta$ is an imaginary number, the solution leads to a contribution that is a wave in the vertical. On the other hand, for $k$ values that do not satisfy this condition, $\beta^2$ is a positive real number. In this case, $\beta$ is chosen as the negative root of $\beta^2$, otherwise the density and pressure perturbation would increase exponentially with height leading to a non-physical solution.

The other variables, horizontal velocity, pressure and density, can be calculated and are

\begin{align}
\tilde{U} & = -(\frac{\delta}{2}+\beta-\frac{g}{c_s^2}) \, \frac{U_0 \tilde{B}}{\alpha_0} \, e^{\beta Z}, \\
\tilde{p} & =  -(\frac{\delta}{2}+\beta-\frac{g}{c_s^2}) \, \frac{\rho_s U_0^2 \tilde{B}}{\alpha_0} \, e^{\beta Z} - \tilde{B} \delta p_s e^{-\frac{\delta}{2} Z}, \\
\tilde{\rho} & = (\delta + \frac{1}{c_s^2}(U_0^2(\beta - \frac{\delta}{2})-g)) \, \frac{\rho_s  \tilde{B}}{\alpha_0} \, e^{\beta Z} - \tilde{B} \delta \rho_s e^{-\frac{\delta}{2} Z}.
\end{align}

Finally, the solution in the original coordinate system $(x,z)$ is obtained by undoing the Fourier, the Bretherton and the coordinate change transformations.

\begin{table}
\caption{Hydrostatic, non-hydrostatic and potential flow tests settings for the background horizontal velocity ($U_0$), Brunt-V\"ais\"al\"a frequency ($N_0$), half width and height of the hill ($a$ and $h$), horizontal and vertical size of the domain ($L$ and $H$). The hill is located in the centre of the domain in all cases. The model runs up to the dimensionless time $t^*=t U_0 / a$, time at which the numerical and the analytical solutions are compared. \label{011706}}
\centering
\begin{tabular}{lrrr}
\toprule
& Hydrostatic & Non-hydrostatic & Potential flow \\
\midrule
$U_0 \, (ms^{-1})$ & $8.0$ & $15.0$ & $15.0$ \\
$N_0 \, (s^{-1})$ & $0.02$ & $0.02$ & $0.02$ \\
$a \, (km)$ & $16.0$ & $0.5$ & $0.1$ \\
$h \, (m)$ & $0.016$ & $0.005$ & $0.01$ \\
$L \, (km)$ & $409.6$ & $38.4$ & $2.56$ \\
$H \, (km)$ & $30.0$ & $30.0$ & $4.0$ \\
$t^*$ & $120.0$ & $90.0$ & $60.0$ \\ 
\bottomrule
\end{tabular}
\end{table}

\begin{table}
\caption{The horizontal and vertical resolution ($dx$ and $dz$), time step ($dt$), the number horizontal and vertical grid points ($N_x$ and $N_z$) and number of time steps ($N_t$), for the hydrostatic ($H$), non-hydrostatic ($N$) and potential flow ($P$) tests. \label{021116}}
\centering
\begin{tabular}{crrrrrr}
\toprule
Test & $dx \, (m)$ & $dz \, (m)$ & $dt \, (s)$ & $N_x$ & $N_z$ & $N_t$ \\
\midrule
$H_1$ & $3200.0$ & $100.0$ & $100.0$ & $128$ & $300$ & $2400$\\
$H_2$ & $6400.0$ & $200.0$ & $200.0$ & $64$ & $150$ & $1200$ \\
$H_3$ & $12800.0$ & $400.0$ & $400.0$ & $32$ & $75$ & $600$ \\
\midrule
$N_1$ & $150.0$ & $150.0$ & $3.0$ & $256$ & $200$ & $1000$ \\
$N_2$ & $300.0$ & $300.0$ & $6.0$ & $128$ & $100$ & $500$ \\
$N_3$ & $600.0$ & $600.0$ & $12.0$ & $64$ & $50$ & $250$ \\
\midrule
$P_1$ & $20.0$ & $20.0$ & $0.4$ & $128$ & $200$ & $1000$ \\
$P_2$ & $40.0$ & $40.0$ & $0.8$ & $64$ & $100$ & $500$ \\
$P_3$ & $80.0$ & $80.0$ & $1.6$ & $32$ & $50$ & $250$ \\
\bottomrule
\end{tabular}
\end{table}

\begin{figure*}[t]
\centering
 
\includegraphics[width=6cm]{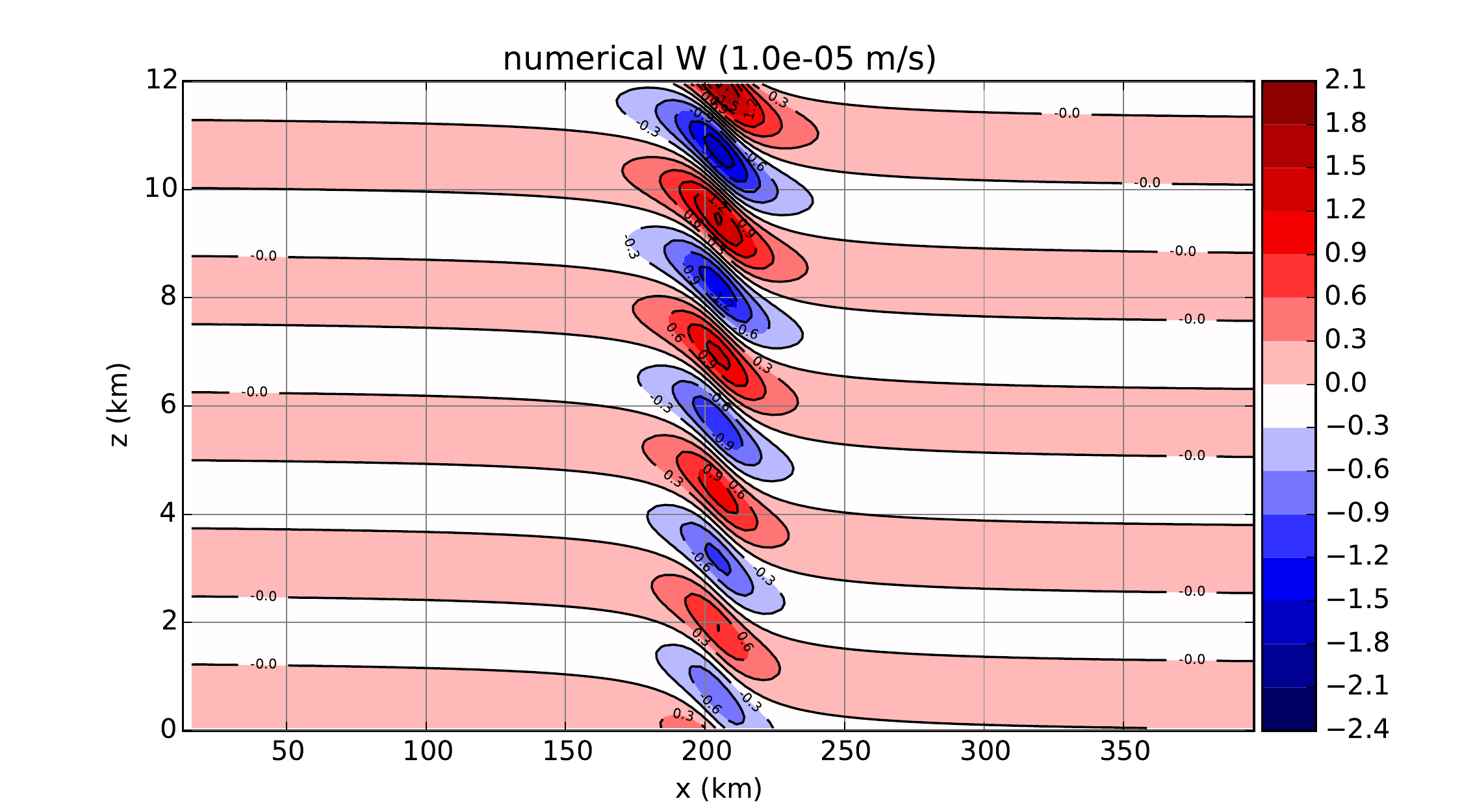}
\includegraphics[width=6cm]{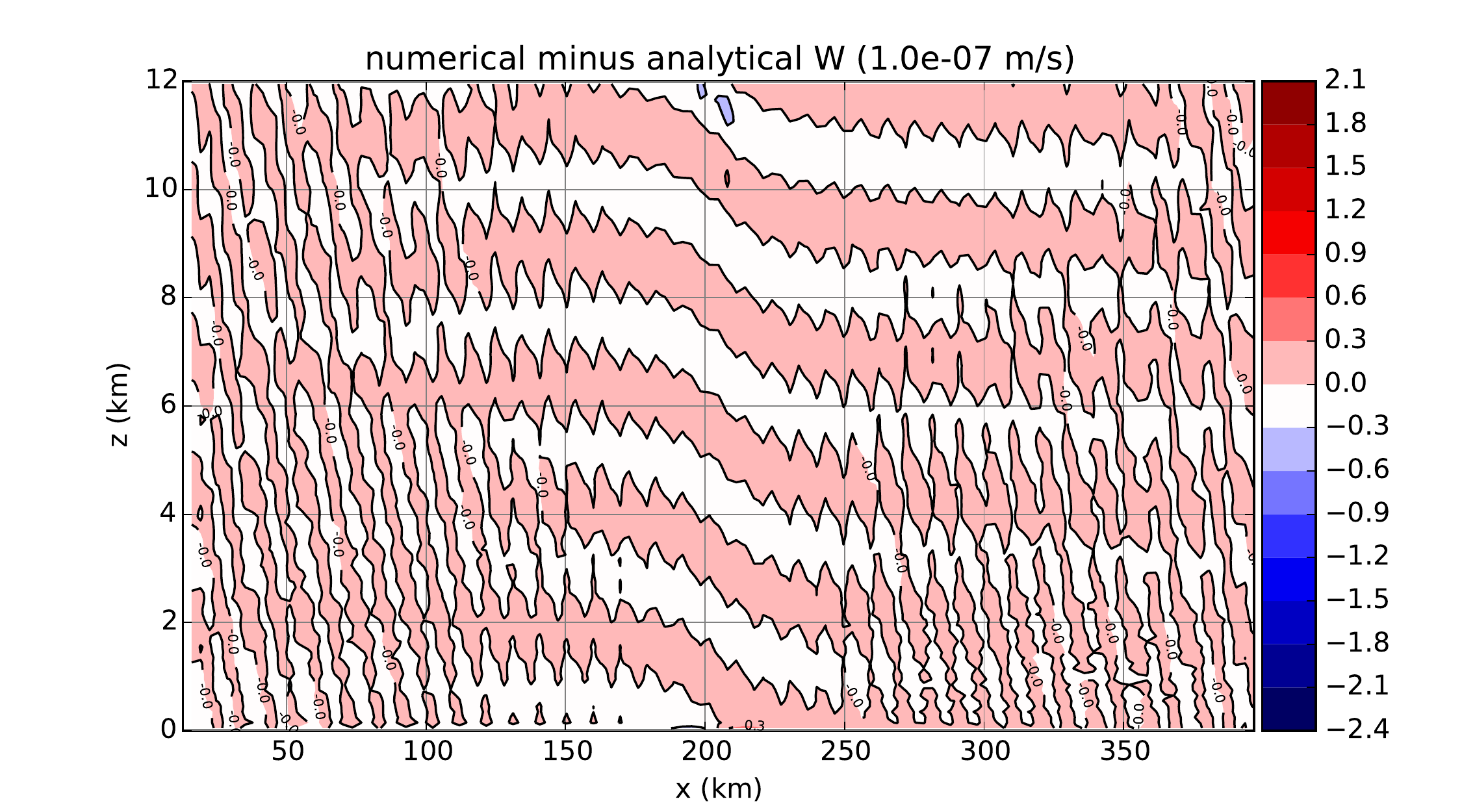}

%\vspace{0.5cm}

\includegraphics[width=6cm]{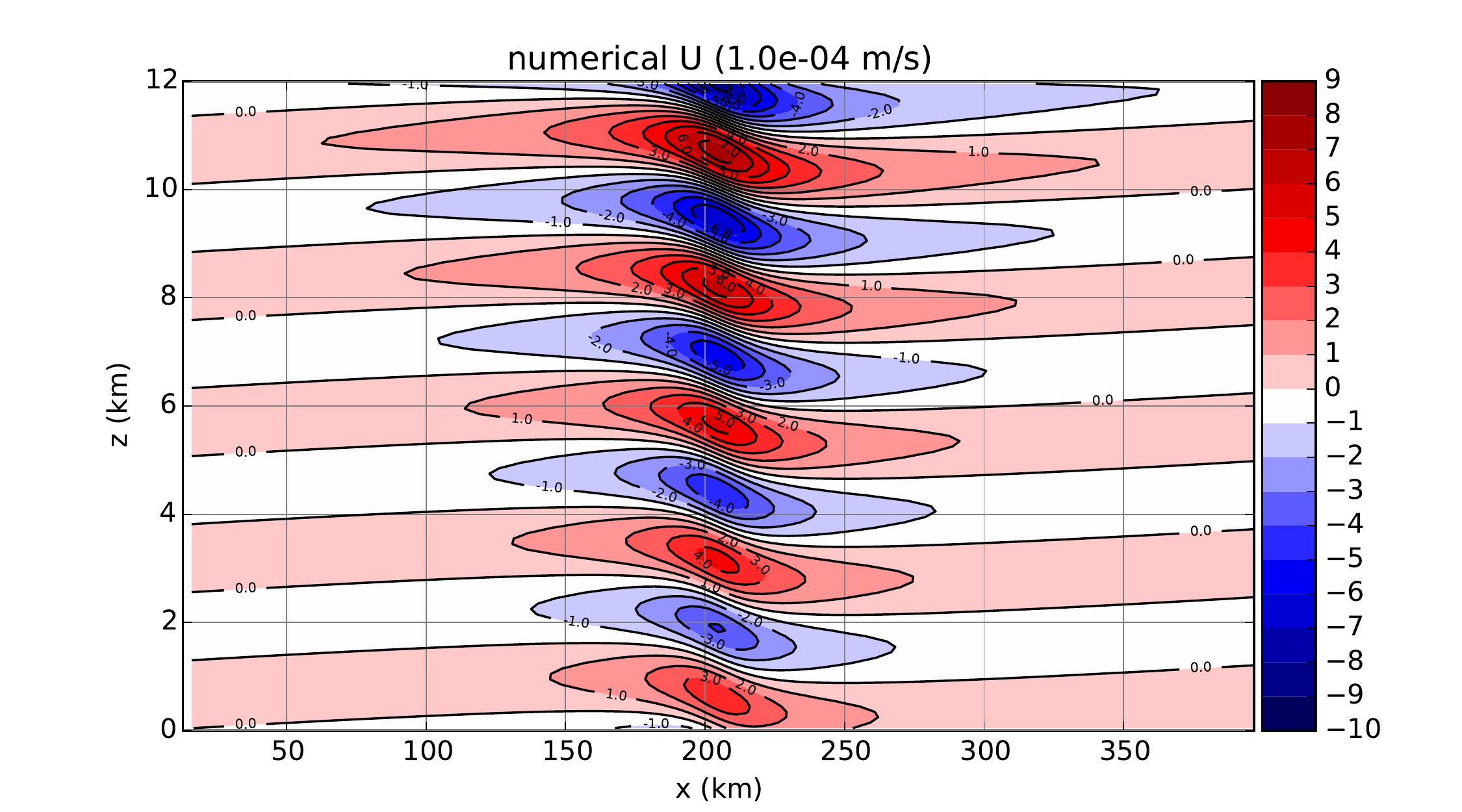}
\includegraphics[width=6cm]{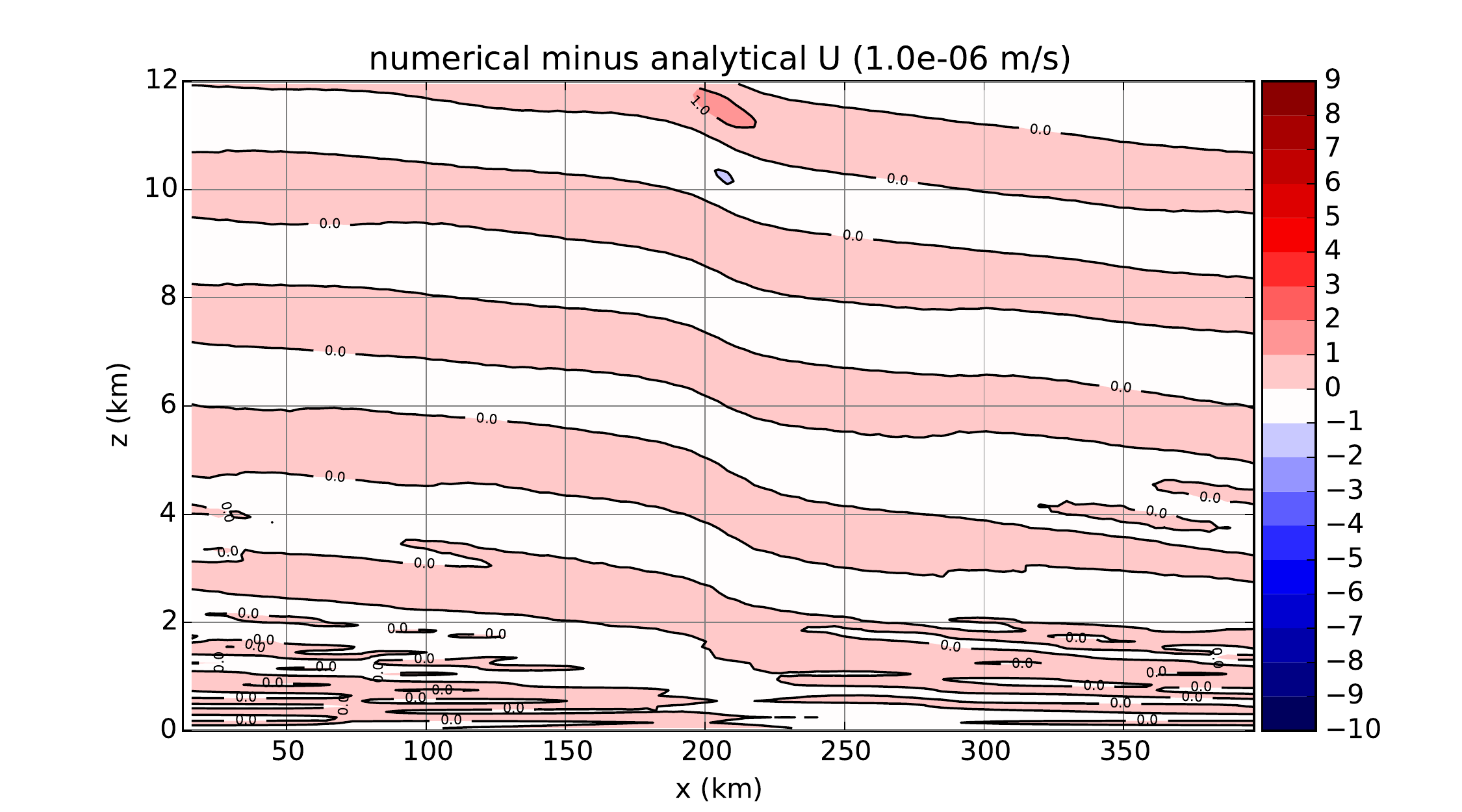}

%\vspace{0.5cm}

\includegraphics[width=6cm]{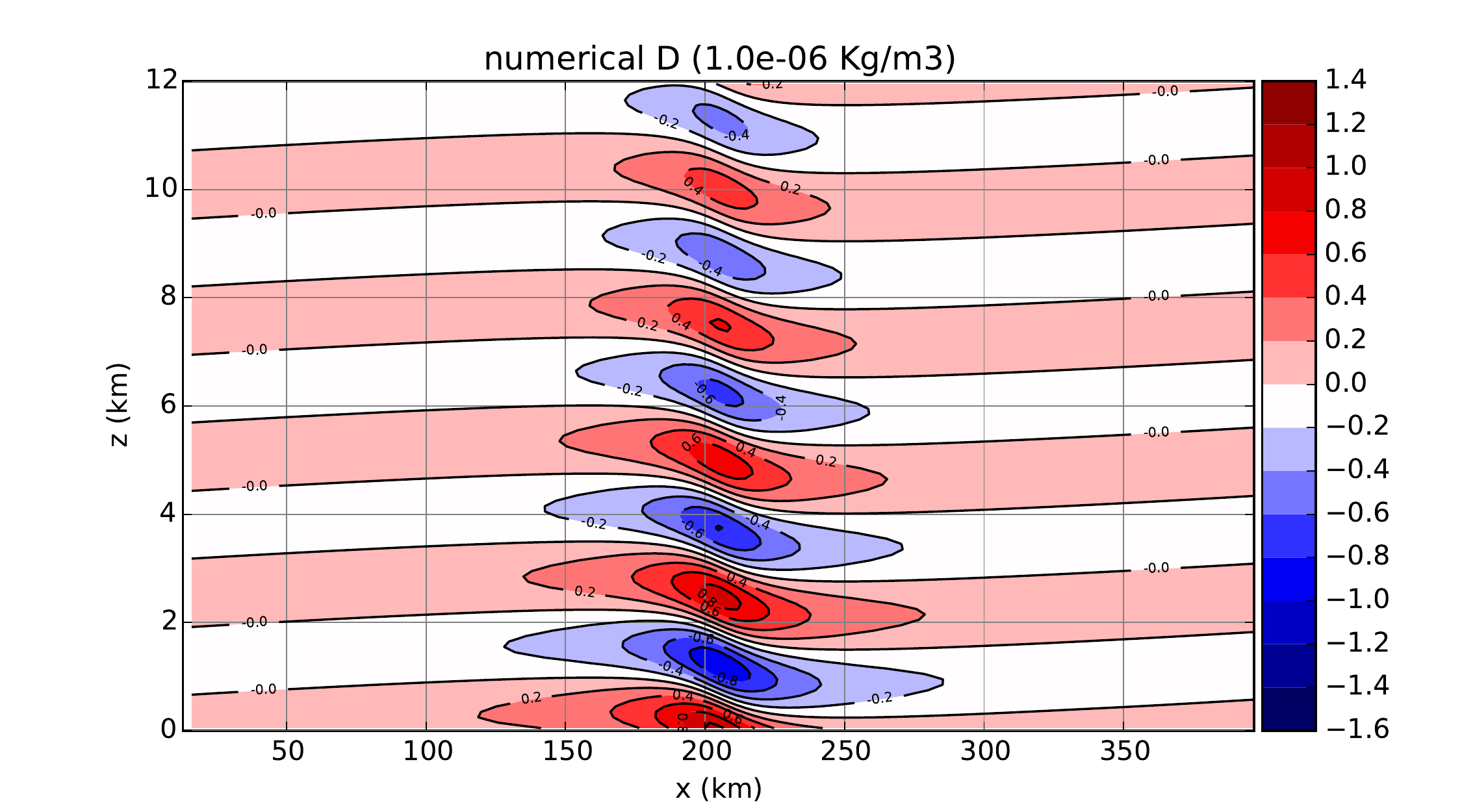}
\includegraphics[width=6cm]{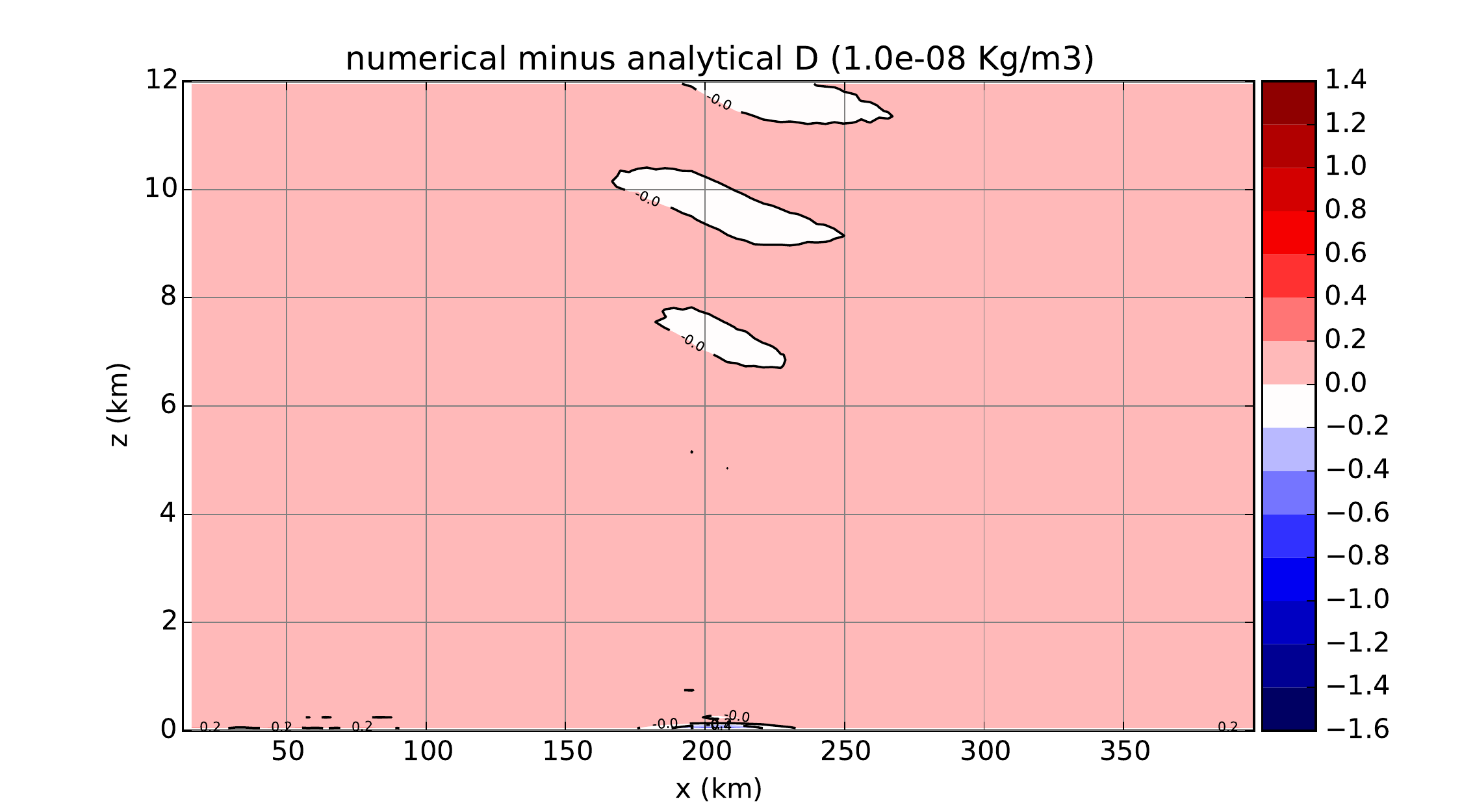}

%\vspace{0.5cm}

\includegraphics[width=6cm]{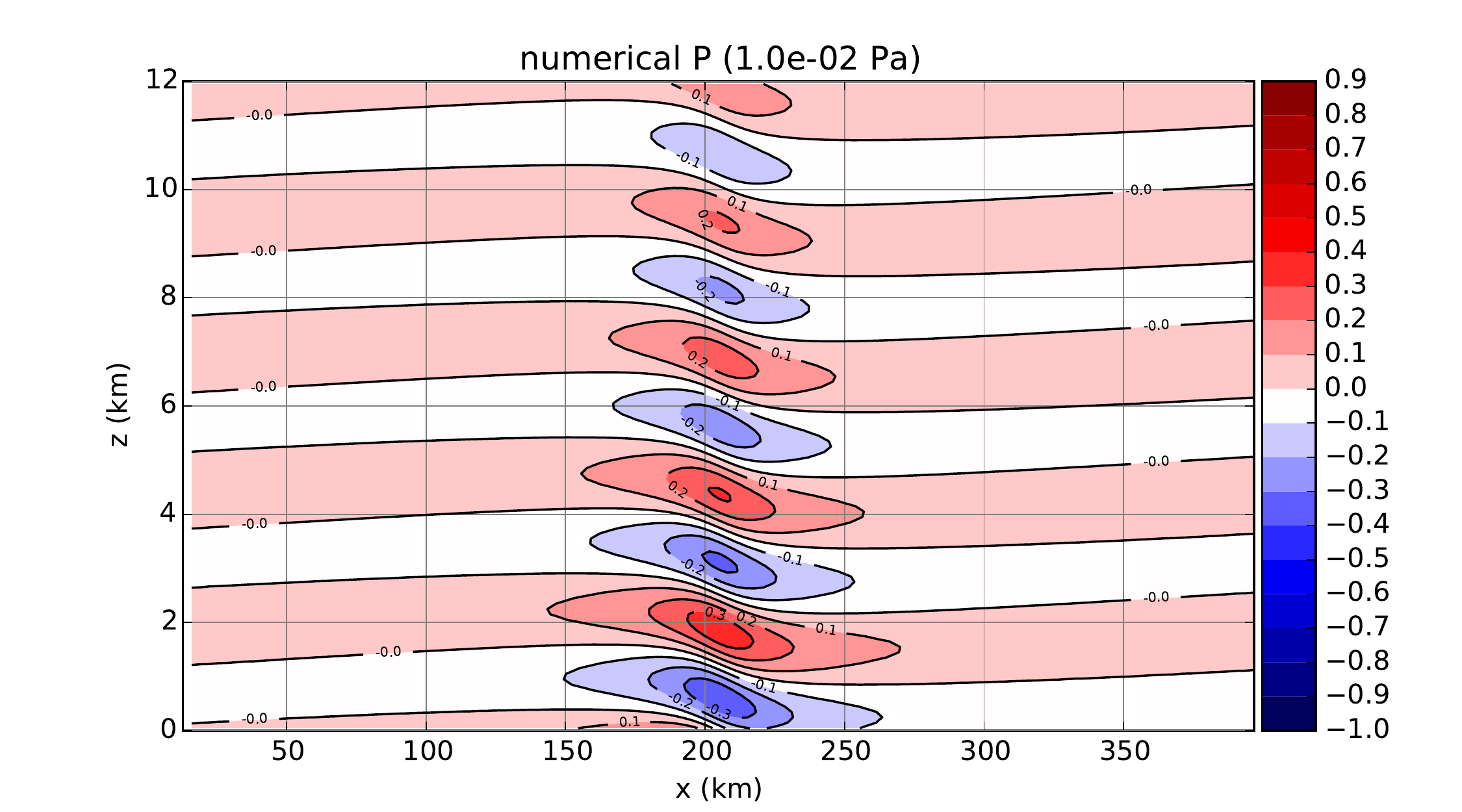}
\includegraphics[width=6cm]{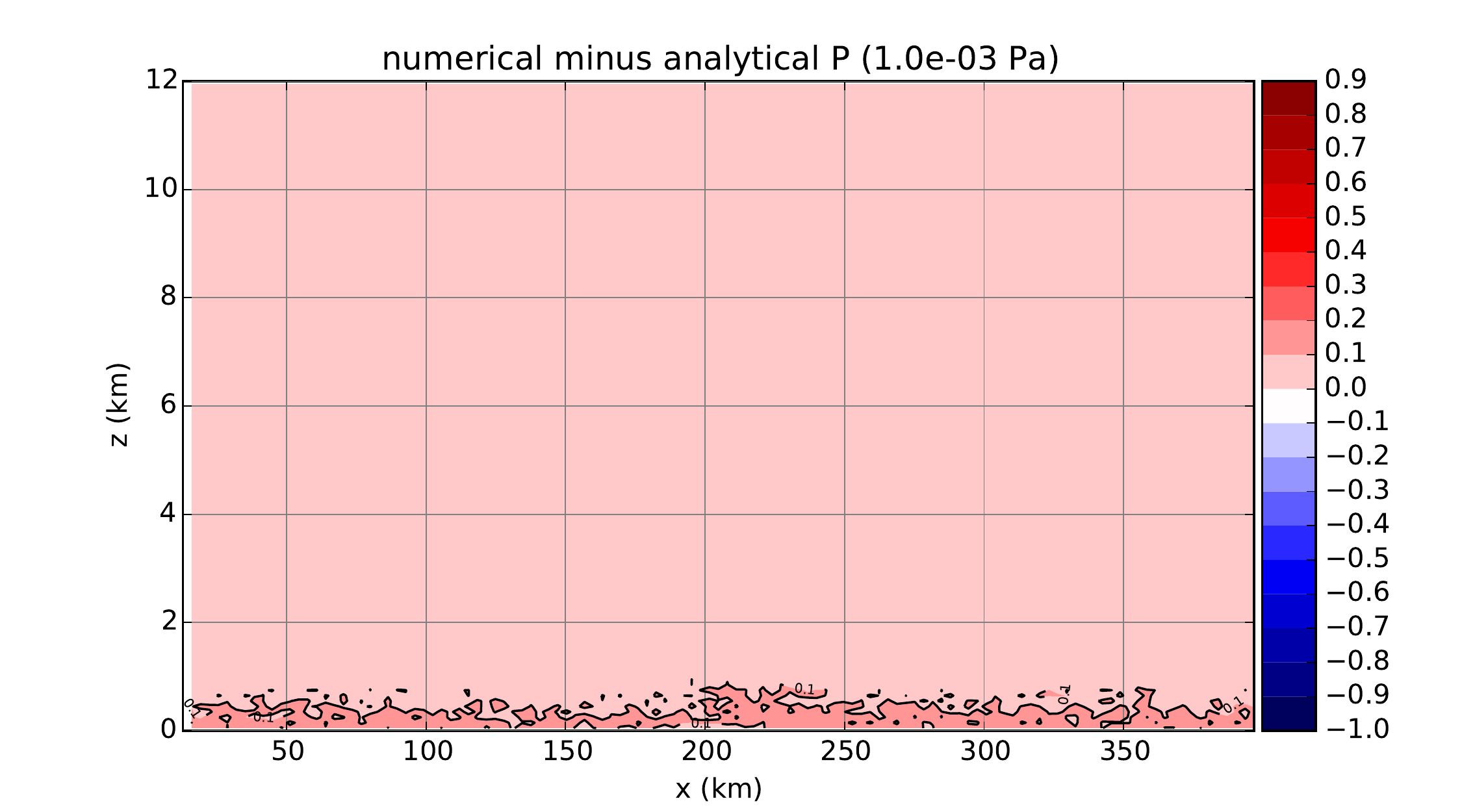}

\caption{Hydrostatic test at the higher resolution ($H_1$): vertical velocity ($W$), horizontal velocity ($U$), density ($D$) and pressure ($P$) perturbations. On the left column it is plotted the numerical solution, and on the right the difference between the numerical and the analytical solutions. \label{0211401}}
\end{figure*}

\begin{figure*}[t]
\centering

\includegraphics[width=6cm]{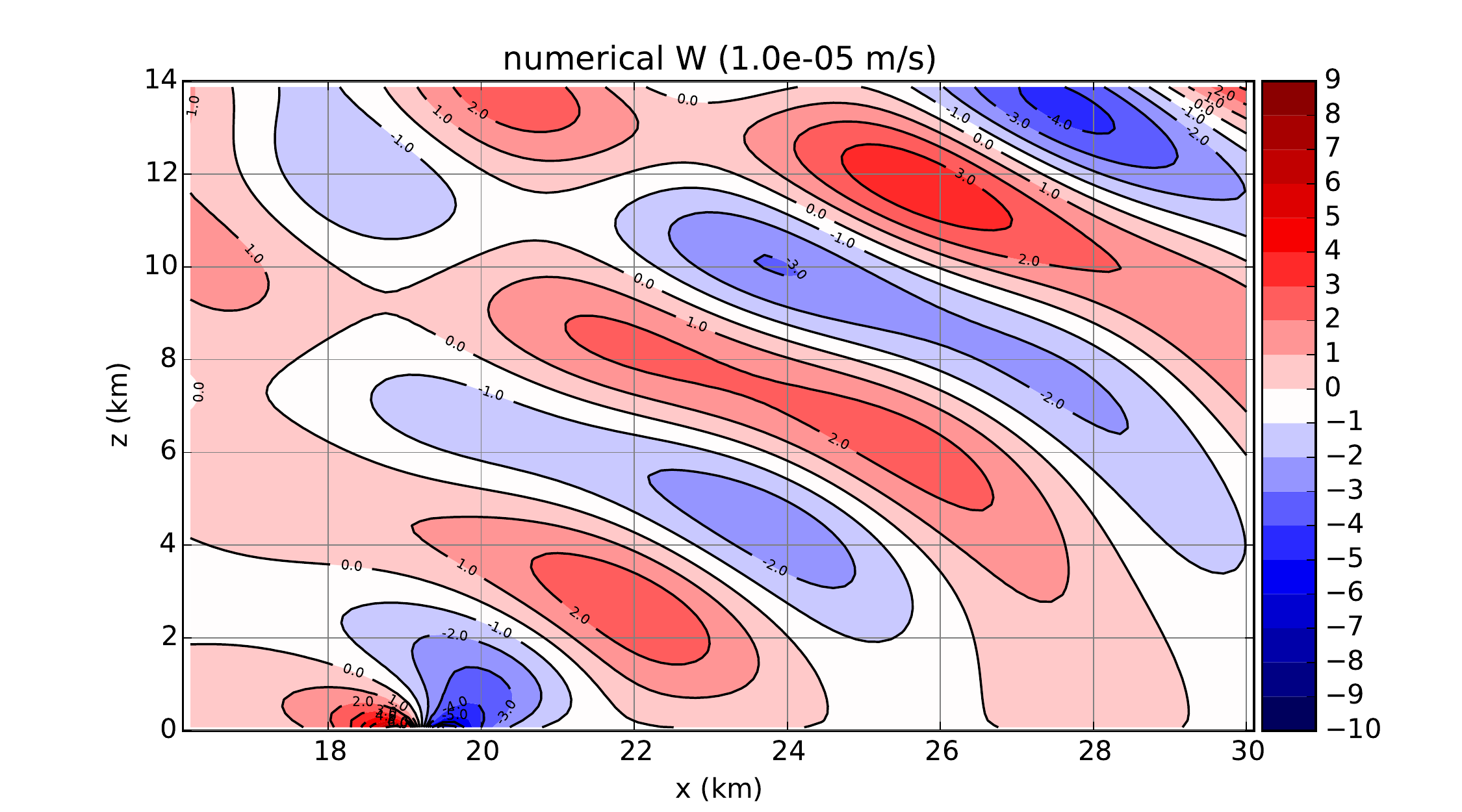}
\includegraphics[width=6cm]	{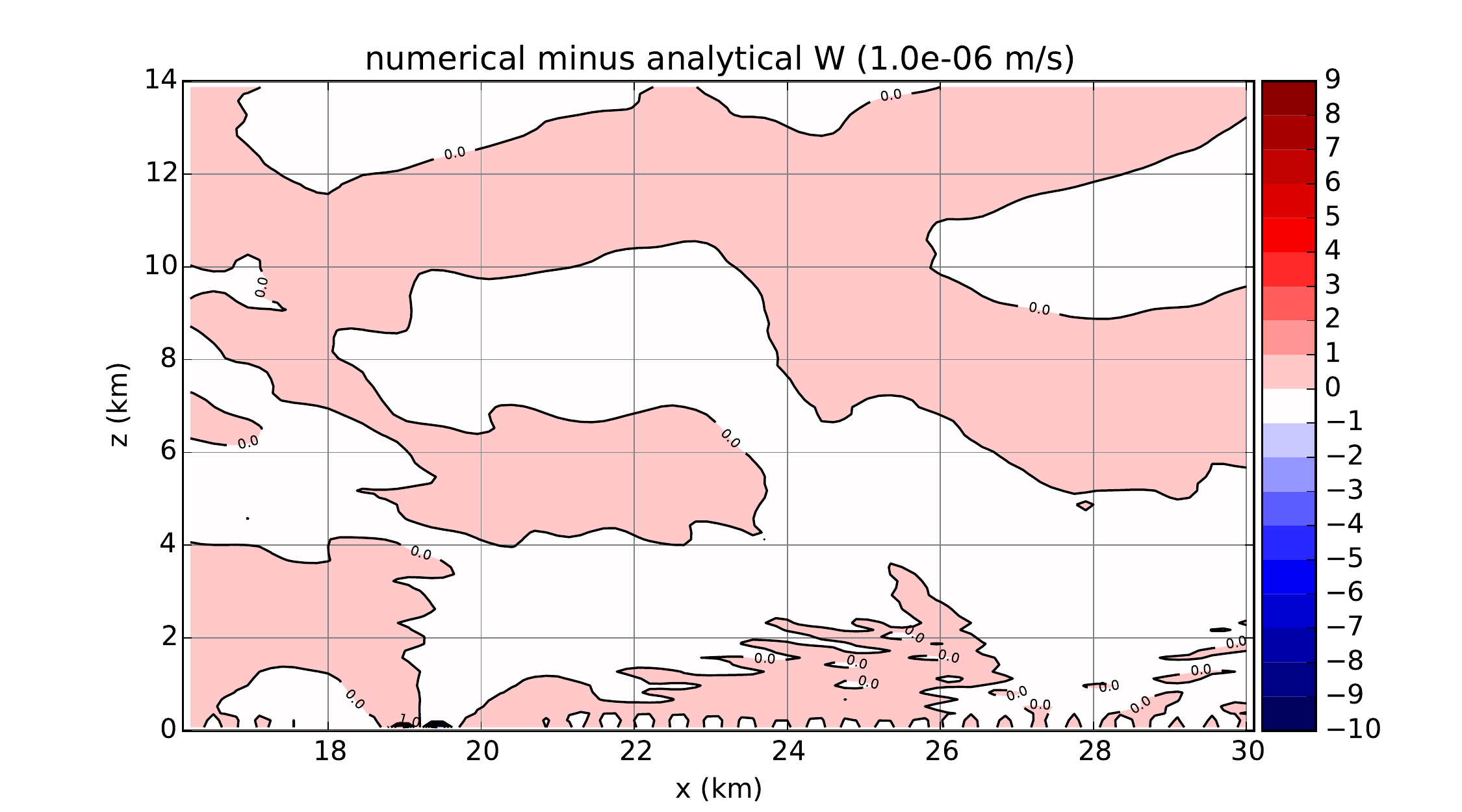}

%\vspace{0.5cm}

\includegraphics[width=6cm]{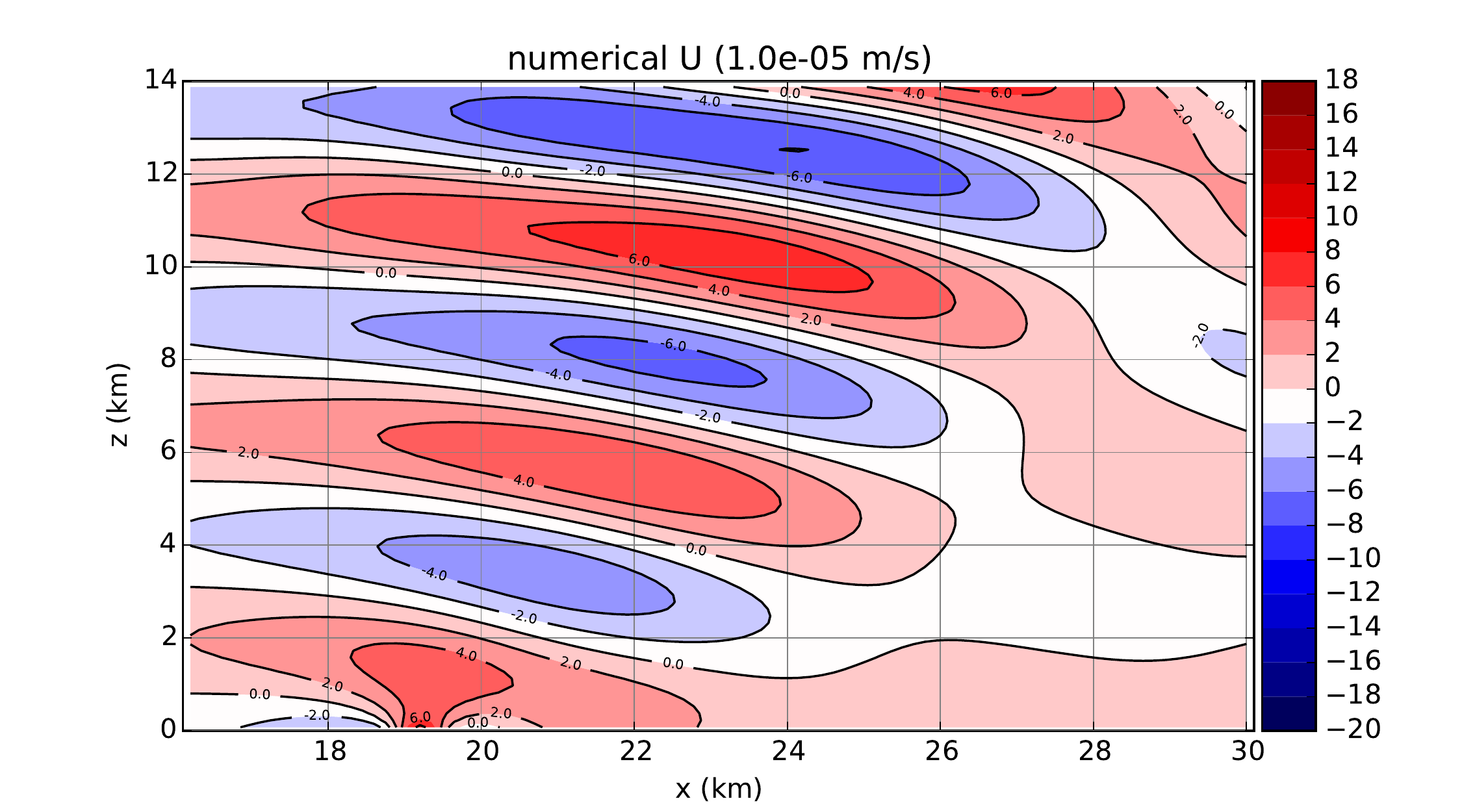}
\includegraphics[width=6cm]{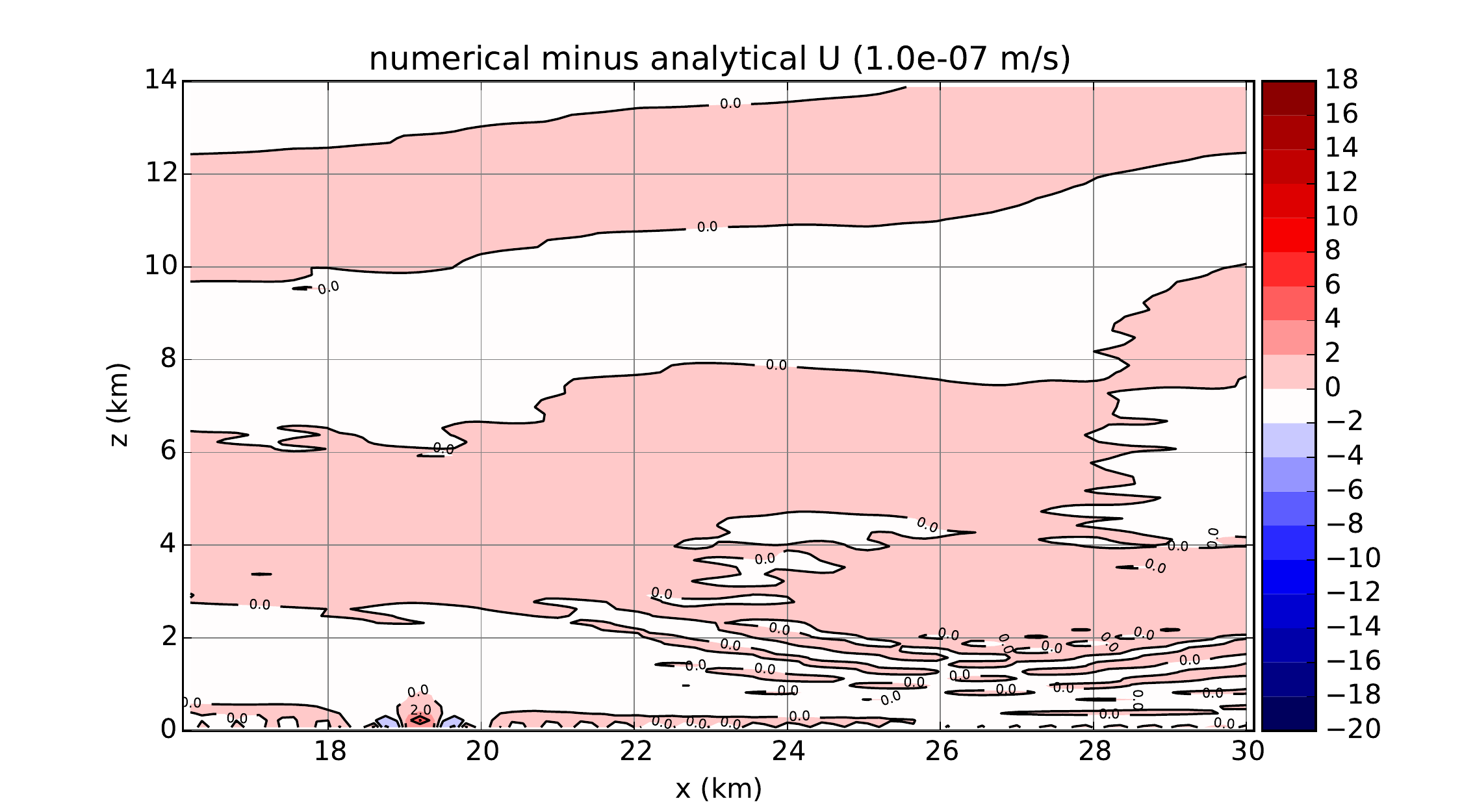}

%\vspace{0.5cm}

\includegraphics[width=6cm]{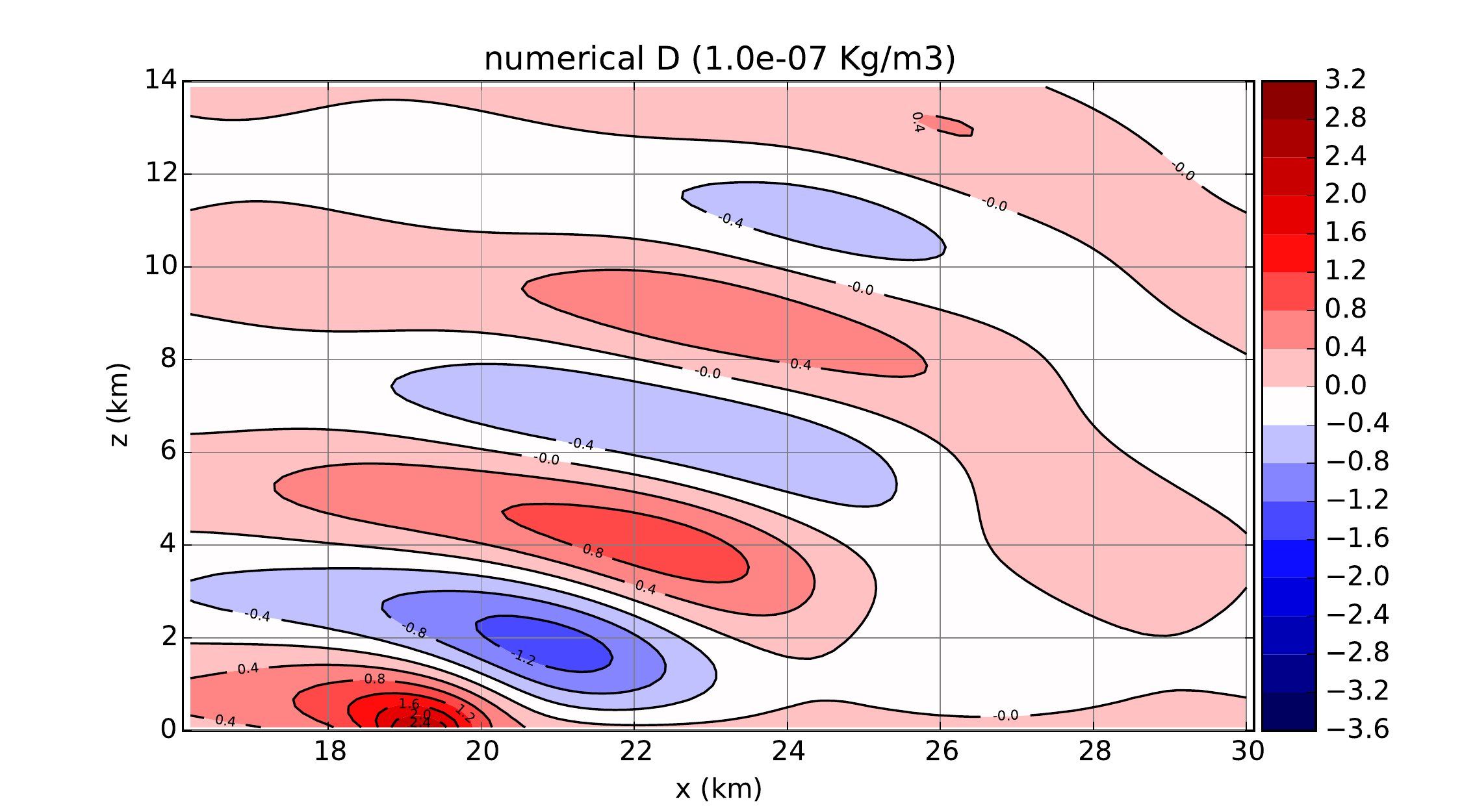}
\includegraphics[width=6cm]{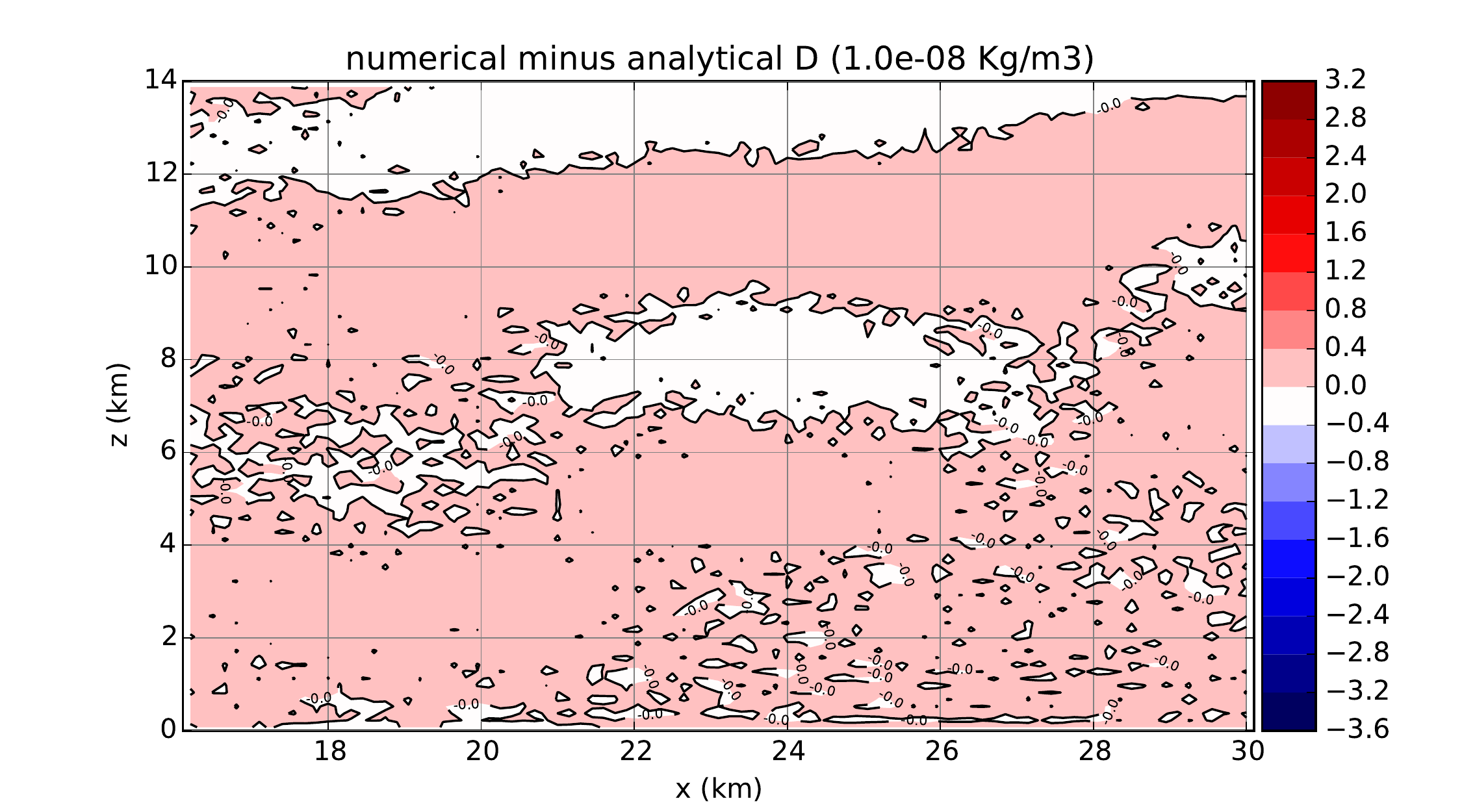}

%\vspace{0.5cm}

\includegraphics[width=6cm]{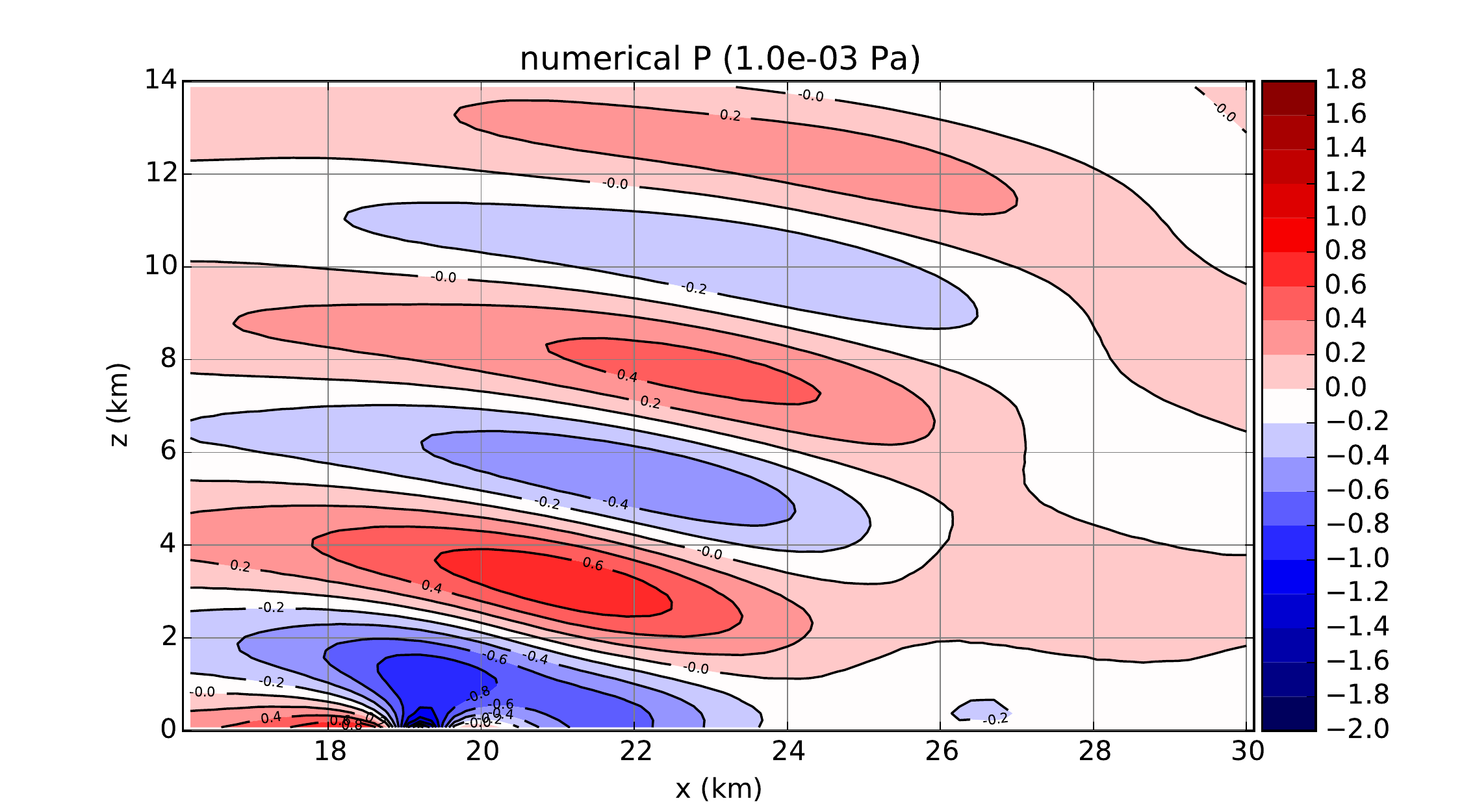}
\includegraphics[width=6cm]{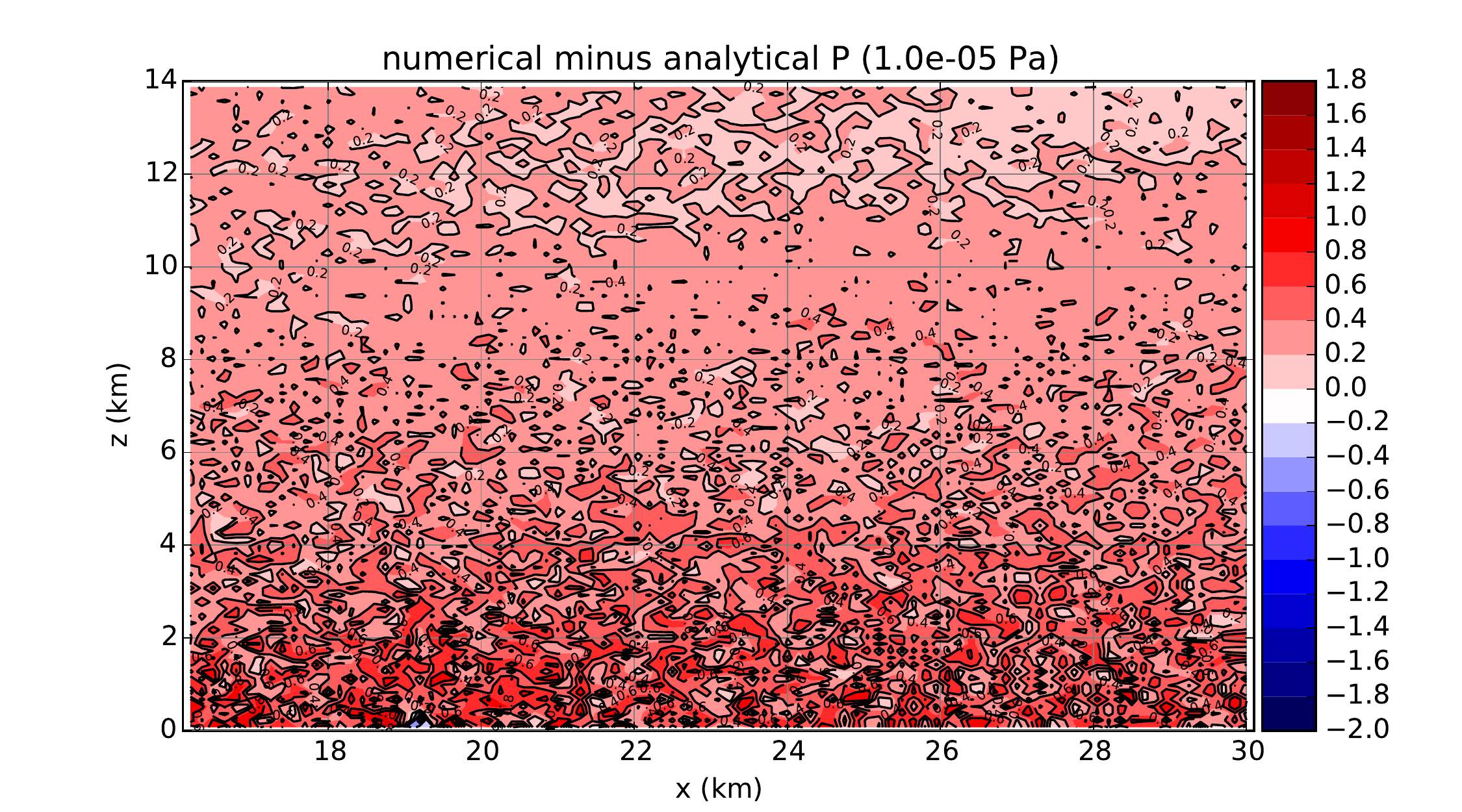}

\caption{Non-hydrostatic test at the higher resolution ($N_1$): vertical velocity ($W$), horizontal velocity ($U$), density ($D$) and pressure ($P$) perturbations. On the left column it is plotted the numerical solution, and on the right the difference between the numerical and the analytical solutions. \label{0211402}}
\end{figure*}

\begin{figure*}[t]
\centering

\includegraphics[width=6cm]{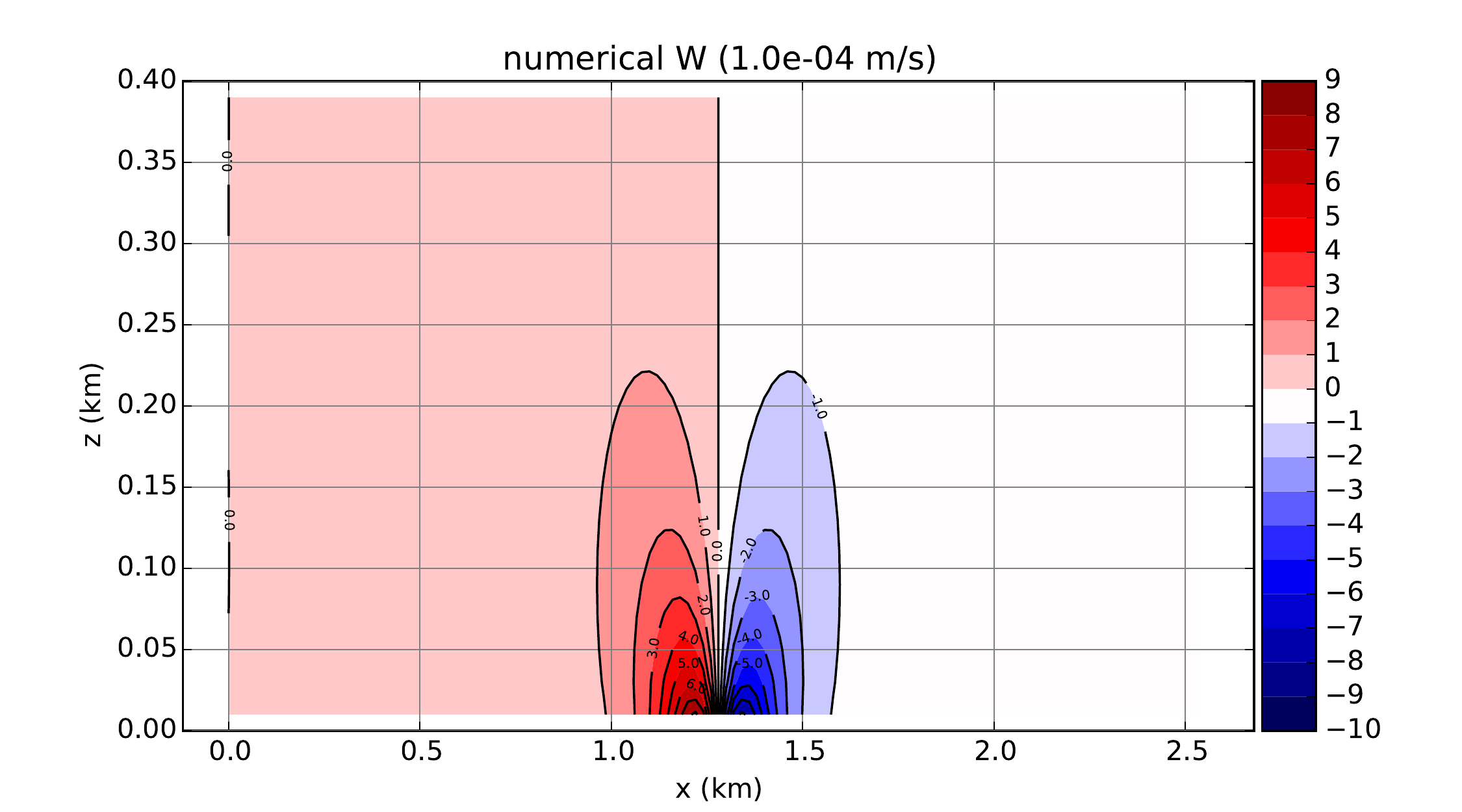}
\includegraphics[width=6cm]	{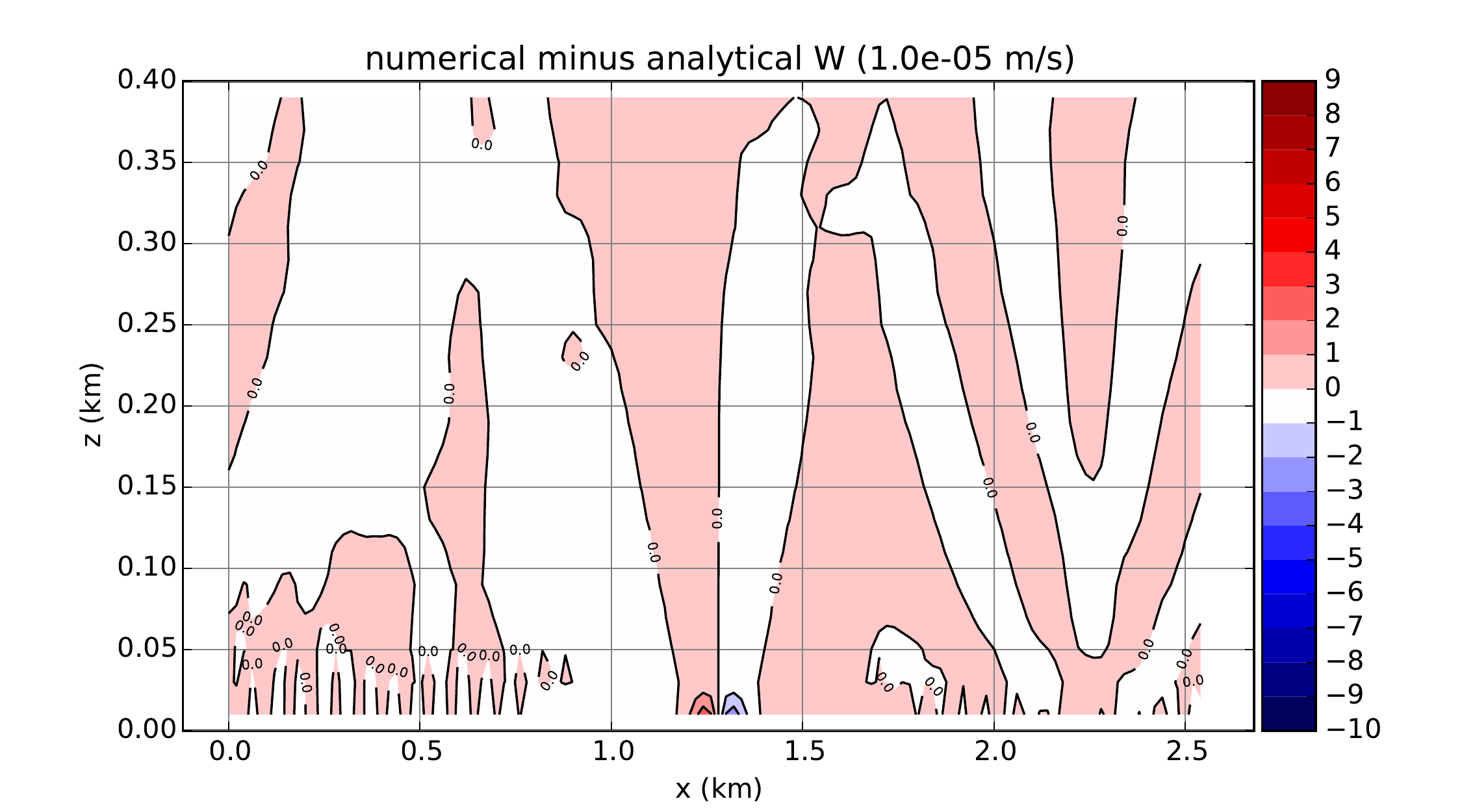}

%\vspace{0.5cm}

\includegraphics[width=6cm]{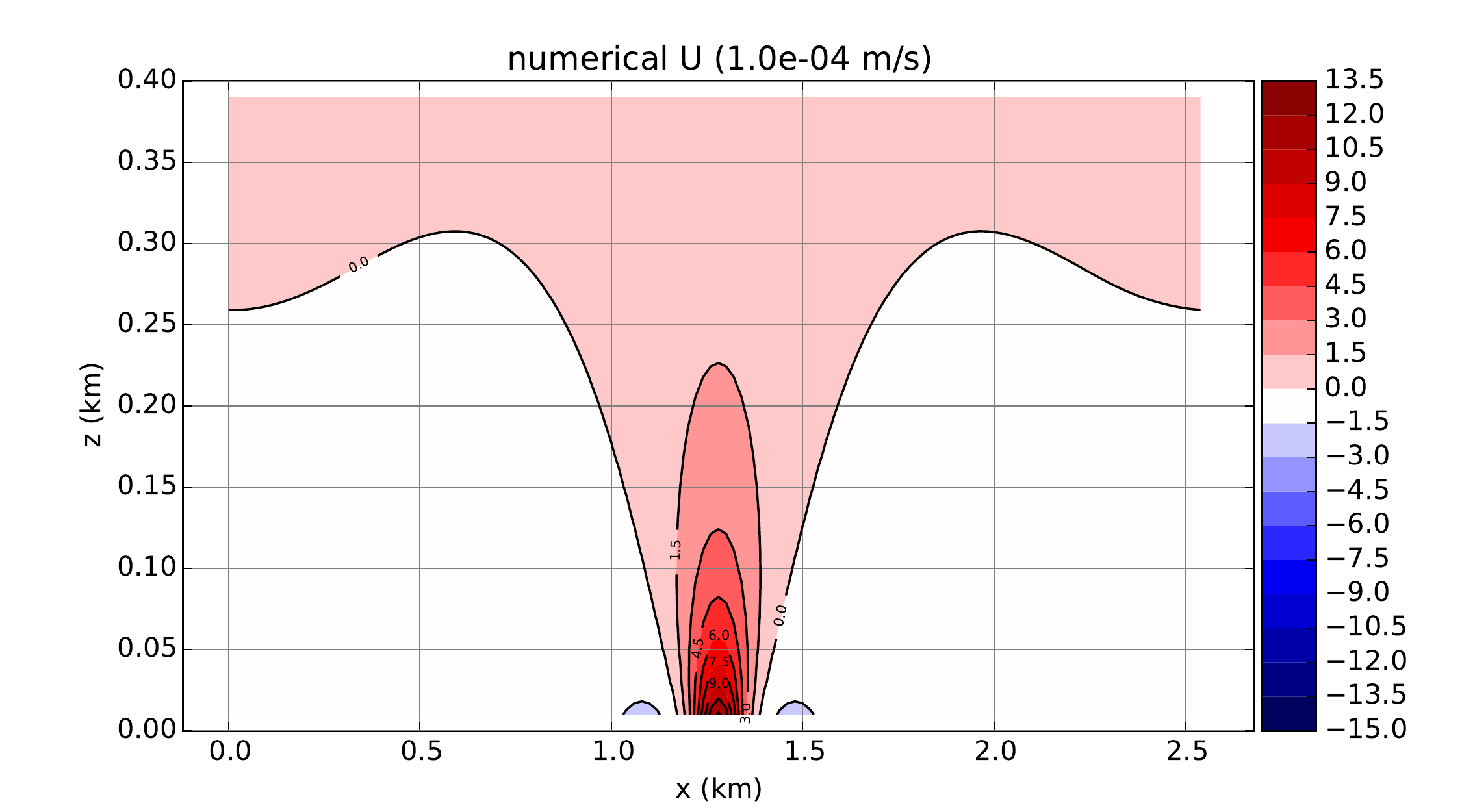}
\includegraphics[width=6cm]{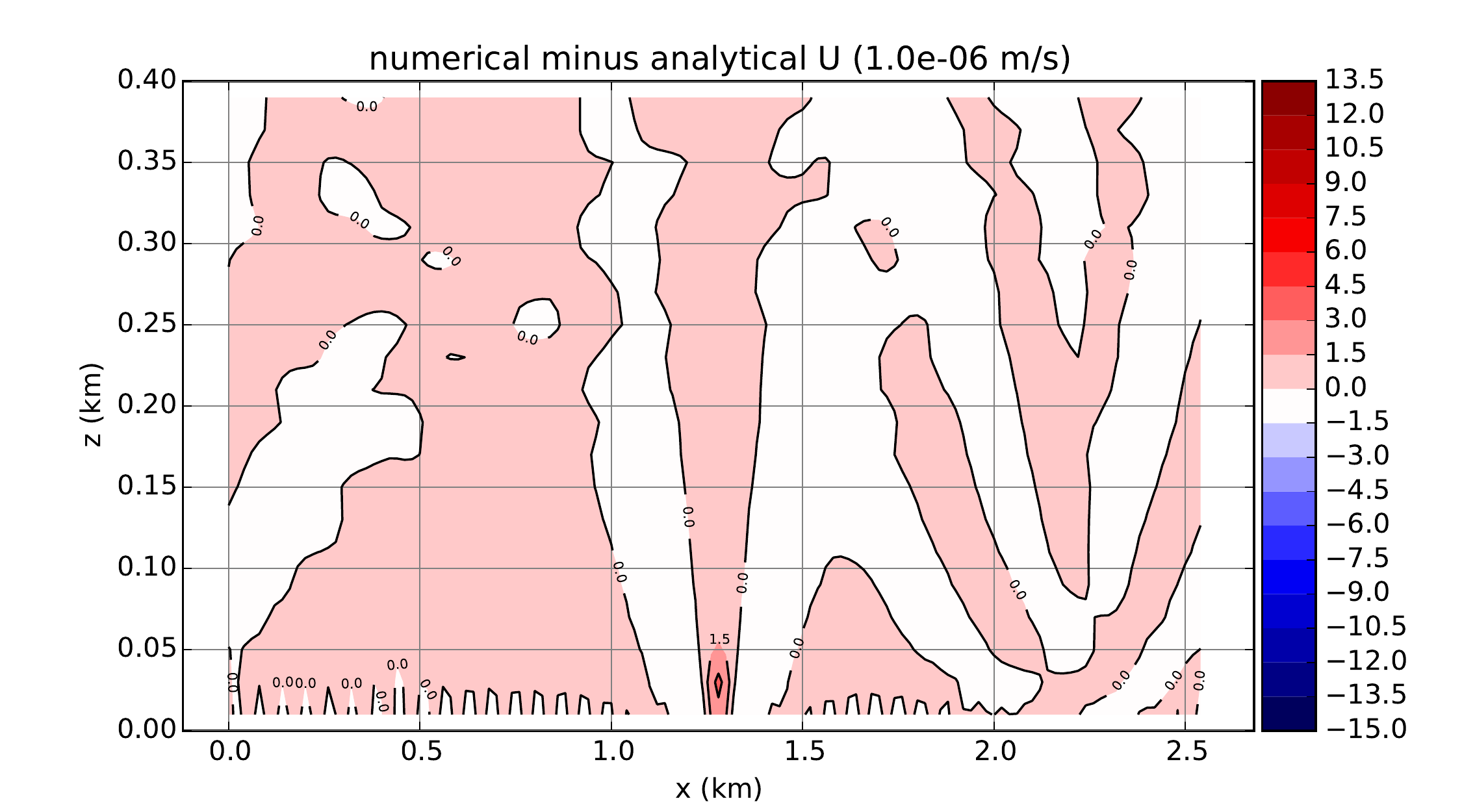}

%\vspace{0.5cm}

\includegraphics[width=6cm]{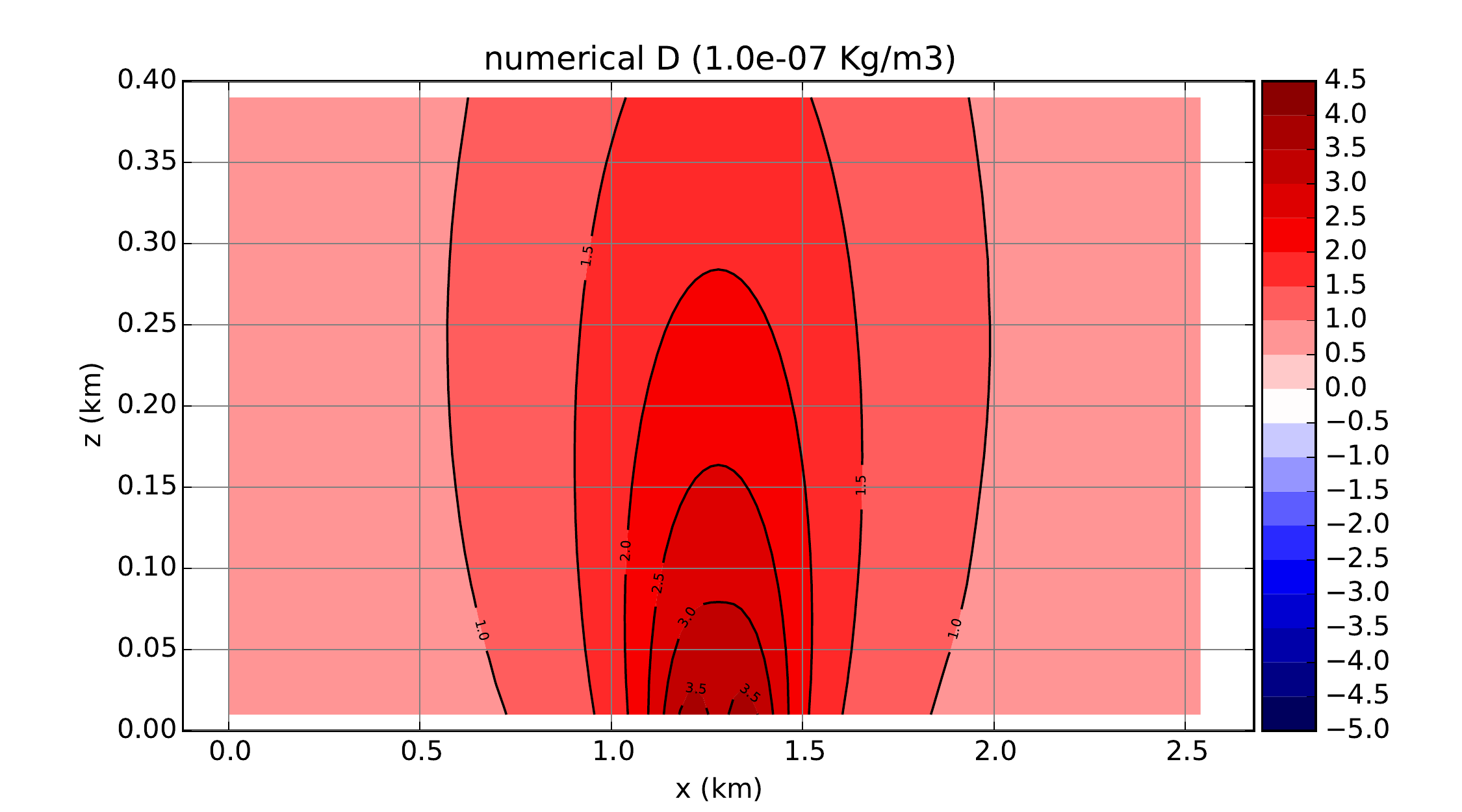}
\includegraphics[width=6cm]{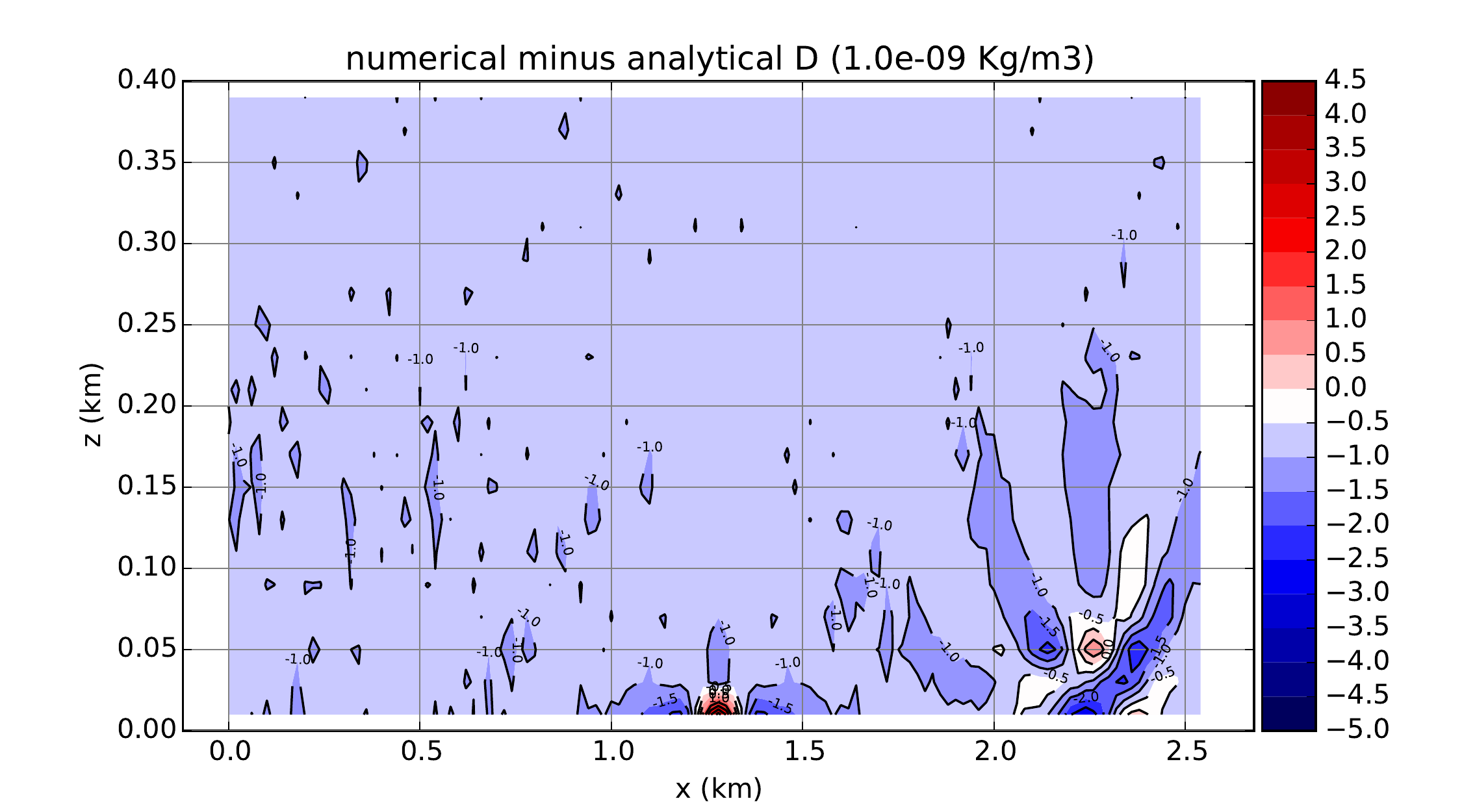}

%\vspace{0.5cm}

\includegraphics[width=6cm]{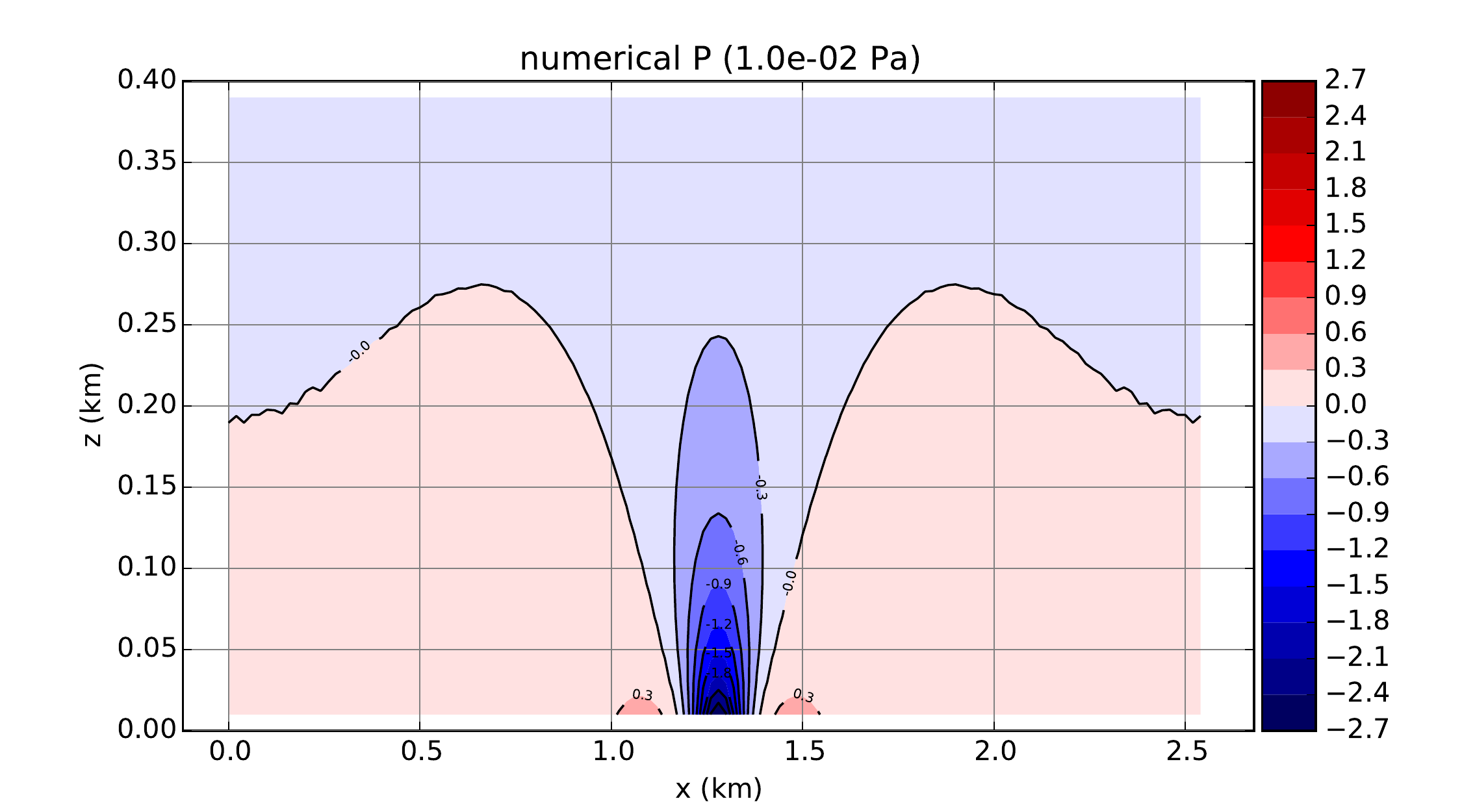}
\includegraphics[width=6cm]{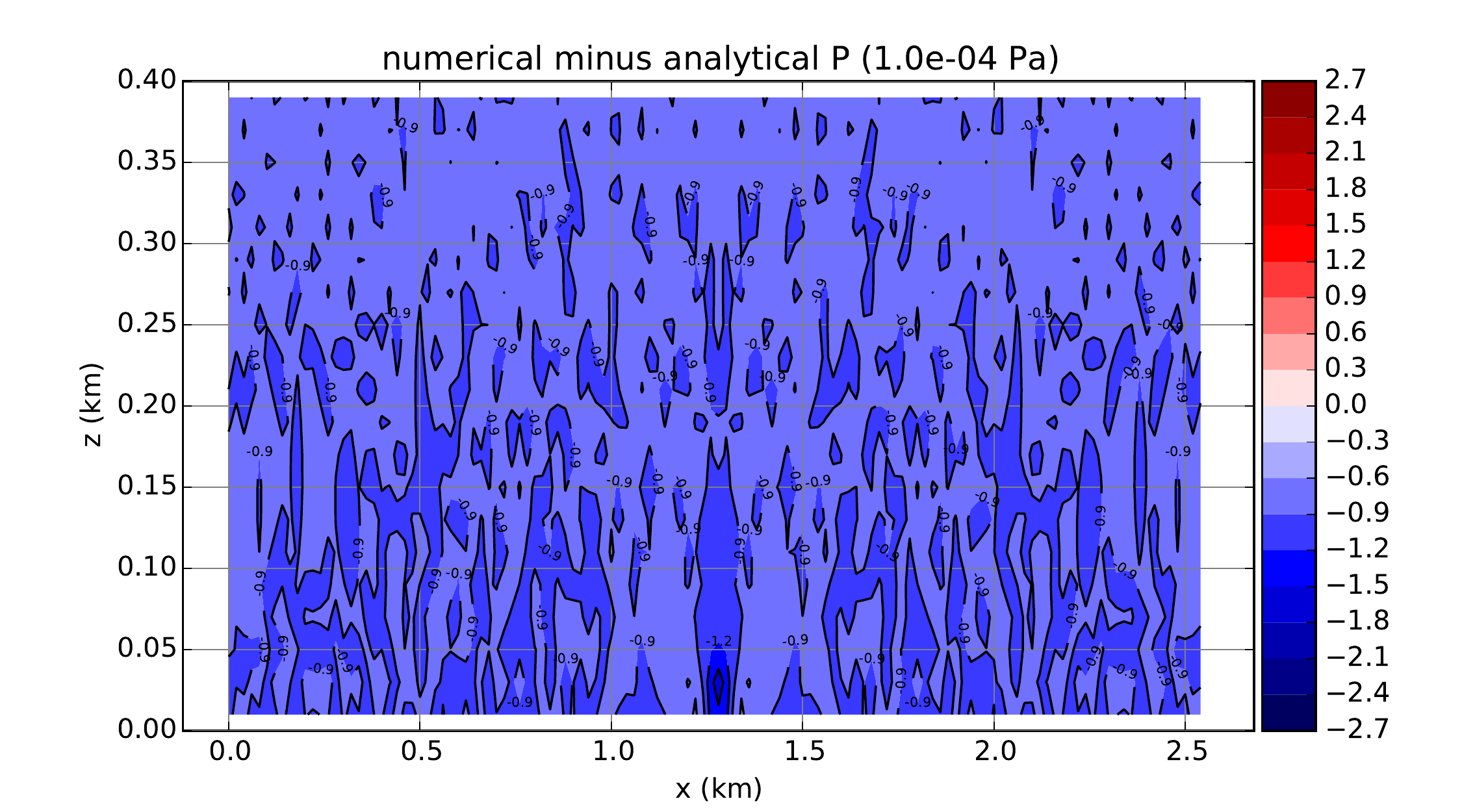}

\caption{Potential flow test at the higher resolution ($P_1$): vertical velocity ($W$), horizontal velocity ($U$), density ($D$) and pressure ($P$) perturbations. On the left column it is plotted the numerical solution, and on the right the difference between the numerical and the analytical solutions. \label{0211403}}
\end{figure*}

\section{Comparison with model simulations}
\label{modelsimulations}

\begin{table}
\caption{Maximum error ($L_\infty$) and standard error ($L_2$) for the hydrostatic ($H$), non-hydrostatic ($N$) and potential flow ($P$) tests at three different resolutions, and variables vertical velocity ($W$), horizontal velocity ($U$), density ($D$) and pressure ($P$). Values are expressed in the international system of units. \label{021109}}
\centering
\begin{tabular}{ccrr}
\toprule
Test & Variable & $L_\infty$ & $L_2$ \\
\midrule
$H_0$ & $W$ & $ 4.475e-08$ & $ 3.673e-09$ \\
$H_1$ & $W$ & $ 1.011e-06$ & $ 1.176e-07$ \\
$H_2$ & $W$ & $ 8.688e-06$ & $ 1.187e-06$ \\
$H_0$ & $U$ & $ 1.234e-06$ & $ 1.789e-07$ \\
$H_1$ & $U$ & $ 2.250e-05$ & $ 3.084e-06$ \\
$H_2$ & $U$ & $ 2.688e-04$ & $ 4.388e-05$ \\
$H_0$ & $D$ & $ 6.026e-09$ & $ 3.597e-10$ \\
$H_1$ & $D$ & $ 3.276e-08$ & $ 3.095e-09$ \\
$H_2$ & $D$ & $ 1.729e-07$ & $ 4.076e-08$ \\
$H_0$ & $P$ & $ 1.088e-04$ & $ 1.725e-05$ \\
$H_1$ & $P$ & $ 1.519e-04$ & $ 2.061e-05$ \\
$H_2$ & $P$ & $ 7.442e-04$ & $ 1.599e-04$ \\
\midrule
$N_0$ & $W$ & $ 4.205e-06$ & $ 8.943e-08$ \\
$N_1$ & $W$ & $ 7.135e-06$ & $ 3.625e-07$ \\
$N_2$ & $W$ & $ 1.408e-05$ & $ 1.441e-06$ \\
$N_0$ & $U$ & $ 6.195e-07$ & $ 4.066e-08$ \\
$N_1$ & $U$ & $ 4.245e-06$ & $ 2.644e-07$ \\
$N_2$ & $U$ & $ 1.126e-05$ & $ 2.475e-06$ \\
$N_0$ & $D$ & $ 3.310e-09$ & $ 8.238e-11$ \\
$N_1$ & $D$ & $ 6.524e-09$ & $ 3.463e-10$ \\
$N_2$ & $D$ & $ 1.858e-08$ & $ 2.587e-09$ \\
$N_0$ & $P$ & $ 1.713e-05$ & $ 1.796e-06$ \\
$N_1$ & $P$ & $ 9.262e-05$ & $ 4.957e-06$ \\
$N_2$ & $P$ & $ 2.353e-04$ & $ 2.501e-05$ \\
\midrule
$P_0$ & $W$ & $ 2.592e-05$ & $ 1.085e-06$ \\
$P_1$ & $W$ & $ 6.223e-05$ & $ 4.750e-06$ \\
$P_2$ & $W$ & $ 9.822e-05$ & $ 1.787e-05$ \\
$P_0$ & $U$ & $ 3.462e-06$ & $ 1.698e-07$ \\
$P_1$ & $U$ & $ 1.209e-05$ & $ 1.399e-06$ \\
$P_2$ & $U$ & $ 7.248e-05$ & $ 1.320e-05$ \\
$P_0$ & $D$ & $ 3.226e-09$ & $ 2.361e-10$ \\
$P_1$ & $D$ & $ 1.276e-08$ & $ 9.008e-10$ \\
$P_2$ & $D$ & $ 3.097e-08$ & $ 3.123e-09$ \\
$P_0$ & $P$ & $ 1.679e-04$ & $ 4.655e-06$ \\
$P_1$ & $P$ & $ 3.761e-04$ & $ 3.038e-05$ \\
$P_2$ & $P$ & $ 1.602e-03$ & $ 2.868e-04$ \\
\bottomrule
\end{tabular}
\end{table}

\begin{table}
\caption{Convergence rate (or slope) and correlation for the $L_\infty$ and $L_2$ errors for the hydrostatic ($H$), non-hydrostatic ($N$) and potential flow ($P$) test and for the variables vertical velocity($W$), horizontal velocity ($U$), density ($D$) and pressure ($P$). \label{021113}}
\centering
\begin{tabular}{cccrr}
\toprule
Test & Variable & Error & Slope & Correlation \\
\midrule
$H$ & $W$ & $L_\infty$ & $3.800$ & $0.994$ \\
$H$ & $W$ & $L_2$ & $4.168$ & $0.993$ \\
$H$ & $U$ & $L_\infty$ & $3.883$ & $0.999$ \\
$H$ & $U$ & $L_2$ & $3.969$ & $1.000$ \\
$H$ & $D$ & $L_\infty$ & $2.421$ & $1.000$ \\
$H$ & $D$ & $L_2$ & $3.412$ & $0.999$ \\
$H$ & $P$ & $L_\infty$ & $1.387$ & $0.936$ \\
$H$ & $P$ & $L_2$ & $1.606$ & $0.900$ \\
\midrule
$N$ & $W$ & $L_\infty$ & $0.872$ & $0.997$ \\
$N$ & $W$ & $L_2$ & $2.005$ & $1.000$ \\
$N$ & $U$ & $L_\infty$ & $2.092$ & $0.983$ \\
$N$ & $U$ & $L_2$ & $2.964$ & $0.999$ \\
$N$ & $D$ & $L_\infty$ & $1.244$ & $0.992$ \\
$N$ & $D$ & $L_2$ & $2.486$ & $0.995$ \\
$N$ & $P$ & $L_\infty$ & $1.890$ & $0.986$ \\
$N$ & $P$ & $L_2$ & $1.900$ & $0.991$ \\
\midrule
$P$ & $W$ & $L_\infty$ & $0.961$ & $0.984$ \\
$P$ & $W$ & $L_2$ & $2.021$ & $1.000$ \\
$P$ & $U$ & $L_\infty$ & $2.194$ & $0.995$ \\
$P$ & $U$ & $L_2$ & $3.140$ & $1.000$ \\
$P$ & $D$ & $L_\infty$ & $1.632$ & $0.992$ \\
$P$ & $D$ & $L_2$ & $1.863$ & $1.000$ \\
$P$ & $P$ & $L_\infty$ & $1.627$ & $0.987$ \\
$P$ & $P$ & $L_2$ & $2.973$ & $0.999$ \\
\bottomrule
\end{tabular}
\end{table}

In this section the analytical linear solution found in the previous section is compared to the numerical solution provided by a non-linear non-hydrostatic numerical model. To this end, we are going to use the model described in \cite{simarro2013}. The model is semi-implict and uses the spectral method in the horizontal discretization. It has a hybrid height-based vertical coordinate. For the experiments presented in this section, it is configured to use fourth order finite differences vertical operators. 

The analytical solution can be used to test the accuracy and convergence of a numerical model. If fact, this was the motivation of this work: to provide an analytical linear solution of the gravitational waves produced in a flow that passes over a hill, with the intention to use it for testing the accuracy of a non-hydrostatic numerical model.

We have selected three different flow regimes for the tests: hydrostatic, non-hydrostatic and a potential flow. The tests configuration are the same as the experiments found in \cite{bubnova1993}. The orography is the Agnesi hill, defined by the half width ($a$) and the height ($h$) of the hill, whereas the flow is determined by the Brunt-V\"ais\"al\"a frequency ($N_0$) and the background horizontal velocity ($U_0$). The values of this parameters are given in Table \ref{011706}. The Agnesi hill, which is located in the centre of the domain, is defined by

\begin{align*}
B(X) = \frac{h \, a^2}{a^2 + X^2}.
\end{align*}

For this type of flow, a constant Brunt-V\"ais\"al\"a frequency and horizontal velocity flow over a localized hill, the type of the waves generated downstream the obstacle depends on the value of the dimensionless parameter $a N_0 / U_0$. For values much greater than one the flow is hydrostatic, for values near one is non-hydrostatic, whereas for values much less than one it becomes a potential flow. For the values given in the Table \ref{011706} it is found that this parameter takes the values $40$, $~0.7$ and $~0.1$, so we expect the flow to be hydrostatic, non-hydrostatic and potential respectively.

The model is initialized with the analytical solution, and it integrates temporally up to $t^* = t U_0 /a$ equal to $120$, $90$ and $60$ for the hydrostatic, non-hydrostatic and potential flow cases respectively, as summarized in Table \ref{011706}. These time lengths are equal to those reported in \cite{bubnova1993}, and are supposed to be big enough to let the model to get a quasi-stationary solution.

The lateral conditions are cyclic, that is, the flow going out from the right boundary is exactly the same as the flow coming in through the left boundary. In the vertical direction there is a sponge layer, from the model top at $H$ to a height equal to $\alpha \, H$, being $\alpha = 0.6$ for all the tests. In this layer the numerical solution is damped towards a predefined solution, which is the analytical solution in the lower domain of the sponge layer, and the background solution towards the top of the sponge layer, being this transition smooth.

The analytical and numerical solutions are then compared at $t^*$. The maximum error ($L_\infty$) and the mean squared error ($L_2$) are calculated for all the variables, vertical and horizontal velocity components, density and pressure. The numerical model has the prognostic variables $r \equiv \log(T)$ and $q \equiv \log(p)$, and therefore, it is necessary to compute the density from them, before doing the comparison. 

The model has been run at three different resolutions for each type of flow, resulting in a total of nine runs. The  vertical and horizontal resolutions, the time step and the number of time steps for each run, are listed in the Table \ref{021116}. 

The hydrostatic test, plotted in Figure \ref{0211401}, effectively consists in a hydrostatic wave that stays over the hill and propagates in the vertical. The difference between the numerical and the analytical solutions are small, compared to the perturbations (later in this section, the maximum and mean squared errors are calculated for all the tests, resolutions and variables). The pattern of the error varies from one variable to other (vertical velocity, for instance, has a horizontal oscillation, whereas pressure error is higher in the lower part of the domain). 

The non-hydrostatic test, plotted in Figure \ref{0211402}, produces a wave that propagates upstream as well as in the vertical. It is remarkable that the errors are quite uniform in all the spatial domain, except for a small area near the hill, where the maximum errors there are located. The pressure, on the other hand, shows a noisy error field, with values that are positive (numerical values higher than the analytical) and bigger in the lower part of the domain.

The potential flow, potted in Figure \ref{0211403}, is a nearly irrotational. The disturbance does not have an oscillatory behaviour, and diminishes exponentially with the height for all the variables. The maximum errors are also located in a reduced area near the hill. The density, moreover, shows a small disturbance propagating within the flow, probably from the beginning of the simulation. 

Every run provide four variables to be compared to their analytical counterparts. Therefore there are $36$ comparisons between numerical and analytical results, with their respective maximum ($L_\infty$) and mean squared ($L_2$) errors, all of them listed in Table \ref{021109}. As it is expected, the errors, both maximum and mean squared, increase when the grid becomes coarser. The analytical and numerical solutions are compared in a window, which is the spatial domain plotted in the Figures \ref{0211401}, \ref{0211402} and \ref{0211403}. Those domains correspond exactly to the figures of the tests mentioned in \cite{bubnova1993}. 

The order of convergence represents how the reduction of the error depends on the increase of the resolution. Theoretically, there is a linear relationship between the logarithm of the spatial resolution and the logarithm of the error, both maximum or mean squared errors. 

For each of the three types of flows (hydrostatic, non-hydrostatic and potential), and for each variable (the vertical and horizontal velocity, density and pressure), we have got three different resolutions and errors (for both $L_\infty$ and $L_2$ norms). Therefore we can perform a linear regression between the logarithms of the vertical resolution and the logarithm of the errors, and test that solutions converge towards the analytical solutions.

In the Table \ref{021113} there is a relation of the slopes and correlations for all the three flows, variables and types of errors. The correlations are remarkable high, only two bellow $0.98$. The convergence rates vary more, and are between $0.872$ (for the maximum error of the vertical velocity in the non-hydrostatic flow) and $4.168$ (for the mean squared error of the vertical velocity in the hydrostatic flow). In all the cases, as expected, the convergence rate of the mean squared error is greater than the respective convergence of the maximum error. 

In the view of the convergence rates and correlations, we can conclude that the numerical solutions converges towards the analytical solutions.

\section{Conclusions}
\label{conclusions}

The aim of this work has been to find an analytical solution of the waves produced in a flow that passes over an orographic feature with small height-width ratio, in order to test the accuracy and convergence properties of the simulations obtained from a numerical model. Because the solution is stationary, the test is better suited to get insight of those aspects of the model that involve the spatial discretization.

For finding the analytical solution we use the covariant formulation of the Euler equations. The coordinate transformation is chosen as the simplest one that makes the lower boundary a coordinate line. After the coordinate transformation, the free slip condition is trivial: the vertical contravariant component of the velocity must be equal to zero. This has the consequence that the orography is not involved in the boundary condition, although it appears in the covariant formulation of the Euler equations in the new coordinates. We could say that the orography has been moved from the boundary condition towards the equations, via a covariant formulation of the Euler equations and a convenient coordinate transformation. 

Once the orography is not involved in the boundary conditions, and it is only present in the equations, the method goes forward through the linearization of the covariant Euler equations. As the orography is supposed to have low height, the linearization procedure also affects the orographic terms, and only those terms that are linear in the orography are retained. Finally, we obtain and solve a linear system that has forcing terms related to the orography. The solution is written, not only for the vertical velocity, but as well for the other variables, horizontal velocity, density and pressure. 

We use a non hydrostatic numerical model for verifying the consistency between the analytical and numerical solutions. The experiments are configured so that the domain is cyclic in the horizontal dimension. In the vertical, a sponge layer is placed in the upper part of the domain. The model is initialized with the analytical solution, and a time integration is performed for a long enough period of time. During the time integration, the model evolves the initial condition towards a numerical quasi-stationary solution. The numerical solution at the end of this time integration is compared to the initial analytical solution. 

The test is configured in three different regimes, hydrostatic and non hydrostatic, and potential flow. It is shown that there is a convergence of the numerical solution towards the analytical solution, when increasing the horizontal and vertical resolutions. The convergence rate varies from one experiment to another, and also from one variable to other. The correlation between the logarithms of the spatial resolution and the errors is very high, showing the consistency of the convergence rates. 

The use of a covariant formulation of the Euler equations has shown to be a successful method for taking into account the orographic forcing in a stationary flow. This method, which up to our knowledge has not been used before in this context, could be extended for other problems. This exploration is left for a future work.

% ===============================================================================

% \ack

% ===============================================================================

% ===============================================================================

\end{document}